\documentclass{aa}
\usepackage{graphicx,txfonts,aas_macros,amssymb}
\usepackage{lipsum}
\begin{document}

  \title{The kinematics of globular clusters systems in the outer halos of the
  Aquarius simulations}


  \author{J. Veljanoski \and A. Helmi}

   \institute{Kapteyn Astronomical Institute, University of Groningen,
              P.O. Box 800, 9700 AV Groningen, The Netherlands \\
              \email{jovan@astro.rug.nl}
             }

   \date{Received \today; Accepted \today}

\abstract {Stellar halos and globular cluster (GC) systems contain
   valuable information regarding the assembly history of their host
   galaxies. Motivated by the detection of a significant rotation
   signal in the outer halo GC system of M31, we investigate the likelihood 
   of detecting such a rotation signal in projection, using cosmological 
   simulations. To this end we select subsets of tagged particles in the halos 
   of the Aquarius simulations to represent mock GC systems, and analyse their
   kinematics. We find that GC systems can exhibit a non-negligible
   rotation signal provided the associated stellar halo also has a
   net angular momentum. The ability to detect this rotation signal
   is highly dependent on the viewing perspective, and the
   probability of seeing a signal larger than that measured in M31
   ranges from 10\% to 90\% for the different halos in the Aquarius
   suite. High values are found from a perspective such that the
   projected angular momentum of the GC system is within $\lesssim
   40$~deg of the rotation axis determined via the projected positions 
   and line-of-sight velocities of the GCs. Furthermore, the true 3D angular 
   momentum of the outer  stellar halo is relatively well aligned, 
   within 35~deg, with that of the mock GC systems. We argue that the net 
   angular momentum in the mock GC systems arises naturally when the 
   majority of the material is accreted from a preferred direction, 
   namely along the dominant dark matter filament of the large-scale 
   structure that the halos are embedded in. This, together with the 
   favourable edge-on view of M31's disk suggests that it is not a coincidence 
   that a large rotation signal has been measured for its outer halo GC system.}


\maketitle
%

\def\kms{km$\;$s$^{-1}$}
\def\Rproj{R_{\rm proj}}
\def\Msun{${\rm M}_{\odot}$}

\section{Introduction}
\label{sec:intro}

According to the currently preferred $\Lambda$CDM cosmological model of galaxy
formation, massive galaxies are at least partly built by hierarchical
coalescence of smaller subgalactic components \citep[e.g.][]{HW1999,Abadi03a,
Abadi03b,BJ2005}. The by-product of such stochastic accumulation of matter are
the diffuse stellar halos surrounding massive galaxies. Even though they contain 
only a small fraction of the total galaxy mass, the far outer halos of galaxies 
are the most likely places in which we are able to directly observe the remnants 
of past accretion events, which -- owing to the long dynamical timescales in 
these regions -- have the form of spatially coherent stellar streams. 
Consequently, the characteristics of these regions provide valuable information 
regarding how galaxies form and evolve through time. Studying the stellar 
populations residing in the outskirts of galaxies, their structure and 
kinematics, is a powerful way to test the predictions of the $\Lambda$CDM 
paradigm in detail, and to gain insight into how galaxies assemble.

Naturally, the most detailed, complete studies of stellar halos to date have 
been conducted in the Local Group using resolved stellar photometry. These 
studies have uncovered a number of stellar streams in the Milky Way 
\citep[e.g.][]{Belokurov06,GrillMair06a,Grillmair06b,Martin14,Bernard14}, which 
are thought to be tidally disrupted dwarf galaxies or star clusters. 
Observations of the Sagittarius dwarf galaxy, which is currently being accreted 
onto our Galaxy, directly shows that the Milky Way's stellar halo is still 
actively evolving \citep[e.g.][]{Ibata94,Majewski03,Koposov12,Slater13}. The Gaia mission will
probe further into the Milky Way halo and provide deeper insight into its
properties. Similarly, wide-field observations of our closest neighbour, M31,
have revealed a highly complex halo dominated by various stellar streams and
overdensities, extending at least out to 150~kpc in projection
\citep[e.g.][]{McConnachie09,Ibata14}. In addition, integrated photometric
surveys have been able to uncover structure in the extended halos of other
nearby galaxies outside the Local Group \cite[e.g.][]{Delgado08,Delgado10}.
Furthermore, numerical simulations have also been able to create lumpy stellar
halos resembling those around Milky Way-sized galaxies that arise as a result
of their hierarchical formation history \citep[e.g.][]{Cooper10,Helmi11}.

The detection of stellar halos is, however, a major observational challenge
owing to the low surface brightness ($\sim~30$~mag~arcsec$^{-2}$) of the stellar
populations that inhabit these regions. An alternative way to study galaxy
halos is through their globular cluster (GC) systems. Because of their high
luminosities, GCs are much easier to observe compared to the underlying
stellar field component. Since GC systems frequently extend to large radii, 
knowledge of their kinematics is of a particular importance. Globular cluster 
motions can provide information on the assembly history of the host galaxy, the 
mass enclosed by the system, the dark matter distribution and the shape of the 
gravitational potential \citep[e.g.][]{Schuberth10,Schuberth12,Strader11}. 
By searching for spatial and dynamical correlations amongst GCs, these objects 
can also be used as tracers of past accretion events 
\citep[e.g.][]{Perrett03,Romanowsky12}.

The first indication that the Milky Way's halo was at least partly formed
through the amalgamation of smaller components came from exploring the
properties of the Galactic GCs \citep{SZ78}. The direct evidence for the
globular cluster population came from exploring the accreting Sagittarius dwarf
galaxy, which is currently found to be donating at least 5 GCs to the Galactic
GC system \citep[e.g.][]{DaCosta95,Ibata95}. Further supporting evidence for the
external origin of the halo GCs in the Milky Way is found by analysing their
ages, metallicities, kinematics, horizontal branch morphologies, luminosities,
and sizes; they are all consistent with the predictions of hierarchical 
formation models 
\citep[e.g.][]{Mackey04,Mackey05,MF09,Forbes10,Dotter11,Keller12}.

Our closest massive neighbour, M31, at a distance of only $\sim
780$~kpc, hosts a rich GC system comprising over 500 confirmed members
\citep{GalletiRBC04}. Recent wide-field surveys showed that this
system is quite extended in nature, with $\sim 90$ GCs having
projected radii larger than 30~kpc, thus probing the outer halo of
this galaxy \citep[e.g.][]{Huxor05, Huxor08,Huxor14,Zinn13}. Indeed,
analysis of the M31's outer halo GC system yielded particularly
interesting results. \citet{Mackey10b} found $\sim 80$\% of the GCs to
preferentially project onto the various stellar overdensities,
suggesting that these objects were brought into the M31 system by
accreted dwarf galaxies, an observation which supports
hierarchical formation models. Recent kinematic studies of the M31
outer halo GC system \citep{Veljanoski13a, Veljanoski14} have found various
velocity correlations or clustering exhibited by GCs that project
along particular stellar debris features. This not only supports the
accretion hypothesis but also highlights the power of using GCs as
tracers of accretion events.

Even more intriguing is the discovery of a high degree of rotation exhibited by
the outer halo GCs out to a radius of $\sim 150$~kpc in projection, which shares
the same direction and rotation axis, and save for the smaller amplitude, as the
GCs that reside in the inner regions of M31 \citep{Veljanoski14}. While it has
been known for quite some time that the clusters dwelling in the disk of this
galaxy rotate quite rapidly \citep[e.g.][]{Perrett02}, such coherent motion is
unexpected for the halo population. For comparison, the halo GCs in the Milky
Way appear to be fully pressure supported \citep[e.g.][]{Harris01}.

Given the compelling evidence that a high fraction of its members have an
accretion origin, it is of interest for galaxy formation studies to
explain what caused the rotation in the M31's halo GC population. One idea that
could explain this observation suggests that dwarf galaxies fell into the
potential of M31 from few preferred directions on the sky, possibly in groups
\citep[e.g.][]{Li08,Libeskind11,Lovell11}. Having support from cosmological
simulations, this idea is particularly attractive because it is also used to
explain the planar structures formed by satellite galaxies around the Milky Way,
M31 and more recently NGC~5128 \citep{Metz07,Ibata13Natur,Tully13,Tully15}.
Interestingly, the plane of satellites uncovered by \citet{Ibata13Natur}, which
contains nearly half of all known satellites around M31, is found to rotate in
the same direction as the halo GC population, although their rotation axes are
offset by $\sim 45^{\circ}$ on the sky. Even though \citet{Veljanoski14} found
no correlation between the dwarf galaxies that are members of the rotating plane
and the outer halo GCs, the parent galaxies of the latter could have had a
similar configuration with correlated angular momenta, much like
the satellites comprising the rotating plane observed today.

Apart from M31, to date only a few attempts have been made to explore the
kinematics of GC systems in spiral galaxies. \citet{Olsen04} studied the GC
kinematics hosted by six spiral galaxies in the nearby Sculptor group. For two
of them, they found the GCs to exhibit significant rotation in the same sense
as the HI gas. For the remaining four, the authors found tentative rotation
signals, which may be influenced by the very low number of clusters.
\citet{Nantais10} analysed the GC system of nearby spiral M81, and found it
to be highly rotating with an amplitude of $108\pm22$~\kms. Their GC sample only
extends to $\sim 10$~kpc in projection, however.

On the other hand, there are numerous studies targeting elliptical
galaxies with varied results. While it is not uncommon to find rotating GC
populations in elliptical galaxies, it is rare to detect such a phenomenon
beyond a few tens of kpc in projection. \citet{Schuberth10} found the blue
GC population around NGC~1399 to show notable rotation out to $\sim 60$~kpc in
projection, while the red GCs are consistent with being fully pressure
supported. On the other hand, in NGC~4636 it is the red GCs that exhibit a
rotation signal out to $\sim 40$~kpc in projection, while the blue populations
show only marginal evidence for such a coherent motion \citep{Schuberth12}.
\citet{Strader11} detected significant rotation of the GCs around M87 out to
at least 55~kpc in projection, while \citet{Blom12} found evidence for rotation
in each of the distinct three subpopulations of GCs around NGC~4365, all located
within 60~kpc in projection. \citet{Pota13} analysed the kinematics of GCs
hosted by 13 early-type galaxies, and found varying results. Typically, the red
GCs were found to exhibit a higher rotation amplitude than their blue
counterparts. These authors also detected a significant rotation signal out to
$\sim 90$~kpc in projection for NGC~1407 and M87.

To gain further physical insight regarding the possible origin of a
rotation signal in galaxy halos, and in particular of M31's halo, we turn to
numerical simulations and explore the kinematics of mock GC systems in galactic
halos created in the Aquarius project \citep{Springel08Aq,Cooper10}. The aim is
to constrain how likely it is for a rotational signal to be detected from a halo
GC population that has an accreted origin, and to what extent one can constrain
the direction of the angular momentum of a stellar halo using the GCs as tracer
objects. The Aquarius galactic halos are excellent study cases because they
share many key properties observed in M31 (see Section~\ref{ss:AqSim}). This
paper is organized as follows: In Section~\ref{s:method} we describe the
Aquarius simulations, how we create mock GC systems around each primary halo,
and how we determine their kinematic properties. In Section~3 we describe the
kinematics of the generated GC systems in a statistical sense, discussing each
Aquarius system individually. Finally, the results of this study are summarized
and discussed in Section~\ref{s:conclusions}.

\section{Methodology}
\label{s:method}

\subsection{The Aquarius simulations}
\label{ss:AqSim}

The Aquarius project is a suite of high resolution simulations that follow the
formation and growth of 6 dark matter halos in a hierarchical $\Lambda$CDM
universe \citep{Springel08Aq}. These halos, dubbed Aq-A to Aq-F, were
selected from the Millenium-II cosmological simulation, and were individually
re-simulated using the parallel Tree-PM {\sc gadget-3} code, an updated version
of {\sc gadget-2} \citep{Springel05}. Each halo comprises more than $10^8$
individual dark matter particles, and has a total mass of
$1-2 \times 10^{12}$~\Msun. Mass values in this range are typically estimated
for the Milky Way and M31. In this paper, we use the `level 2' simulations, the
highest resolution available for all six halos\footnote{In the
subsequent analysis, we only consider halos Aq-A to Aq-E. Nearly the entire
stellar halo of Aq-F was built through a major merger at $z\sim0.3$ making it
unrepresentative of the Milky Way or M31.}.

\begin{figure*}
\centering
\includegraphics[width=\textwidth]{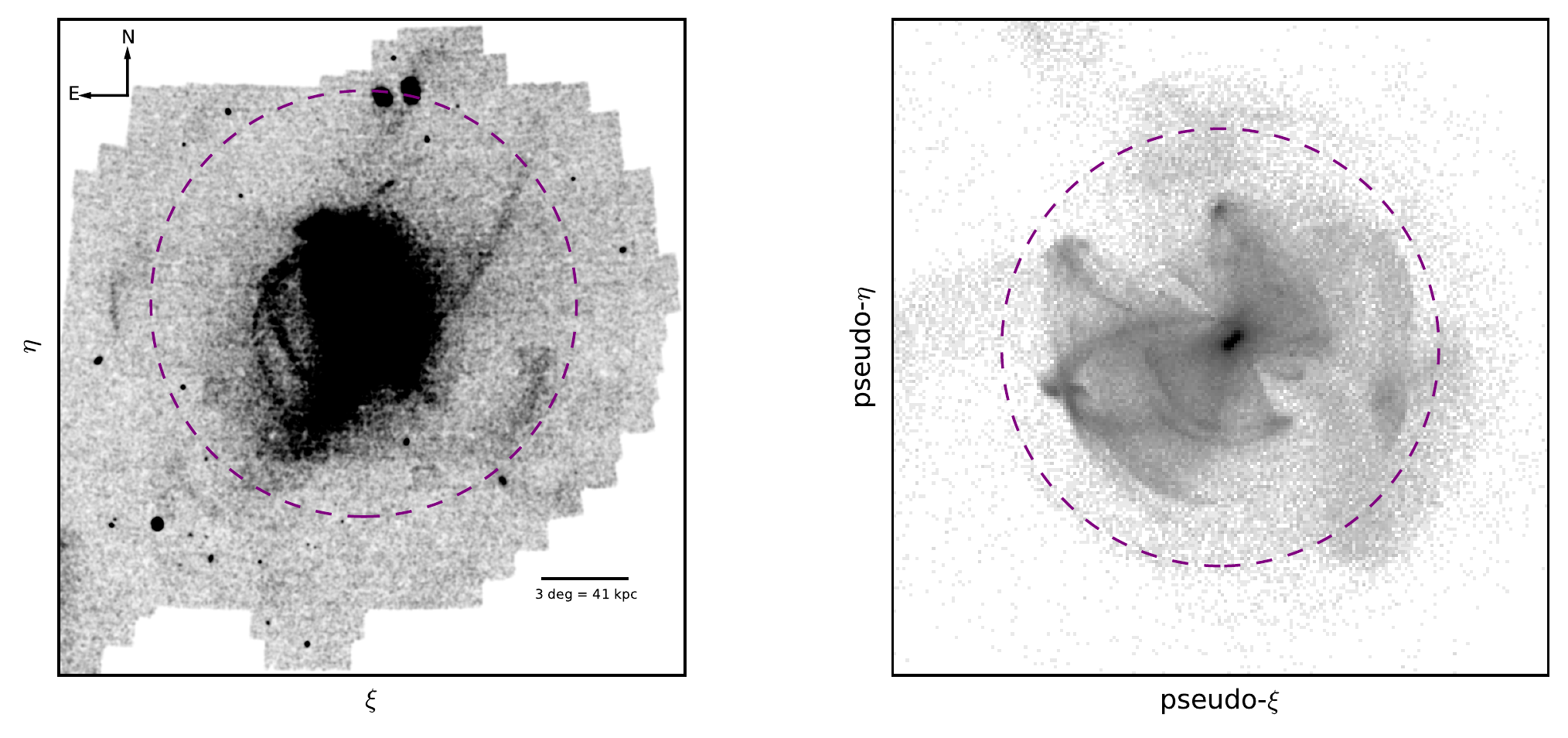}
\caption{Left panel: Metal-poor ([Fe/H]$\lesssim-1.4$) stellar density map
of M31 in standard tangential coordinates. Right panel:  Equivalent
stellar halo of the simulated Aq-C, seen from 1~Mpc away. The displayed
region is 300$\times$300~kpc wide. To give a better sense of scale, a dashed
purple circle with a radius of 100~kpc is drawn in both panels. Similar debris
resulting from accretion events are easily seen in the halos of both M31 and
Aq-C. In the real and the simulated galaxy, the accreting substructures have
anisotropic distributions on the sky. We note that the entire Aq-C stellar halo 
is built from accreted material since the simulations contain no in situ
components.}
\label{f:M31vsAq}
\end{figure*}

By coupling the dynamical information of the dark matter particles to
{\sc galform}, the Durham semi-analytic galaxy formation model \citep{Bower06},
\citet{Cooper10} was able to imprint baryonic content into the Aquarius halos.
To do this, at each output of the dark matter simulation, \citet{Cooper10}
assigns stellar properties such as ages, luminosities, masses and metallicities
to 1\% of the most bound dark matter particles in those halos for which the
semi-analytic model predicts that stars will be formed. The 1\% fraction is
chosen so that the models can reproduce the observed distributions of half-mass
radii, as well as the observed profiles and velocity dispersions of the
surviving satellites. This method, with which $4-6 \times 10^5$ particles are
tagged, is very suitable for tracing the build-up and evolution of accreted
stellar halos without needing to include gas physics. In addition, the
post-processing done in this way is considerably less computationally expensive
than running a hydrodynamical simulation on the same scale; the semi-analytic 
model takes short time to execute and it can be run multiple times in order to 
determine the most realistic values for the free parameters.

Given that the dynamics in these simulations are collisionless, it is important
to note that these models do not track the formation and evolution of in situ 
stellar populations. For example, our sample of Aquarius halos are likely to 
host a massive disk, since no major merger has occurred in the past several 
Gyrs \citep[see e.g.][]{Scannapieco2009}. The existence of such a disk does mean 
that a galaxy is likely to have a well-defined angular momentum, at least in the 
inner regions. However, the orientation of the angular momentum of the disk is 
often correlated with that of the dark matter halo \citep{Bett10}. In addition, 
the existence of a massive disk is likely to influence the orbits and the 
disruption times of the infalling satellites. Nonetheless, we do not expect this 
to have any qualitative impact on our analysis.

The Aquarius simulations are thus well suited for achieving our goal, which is
to study the kinematics of outer halo GC systems that  primarily have an 
external origin, hosted by galaxies analogous to M31. Figure~\ref{f:M31vsAq} 
presents important similarities between M31's stellar halo, shown in the left 
panel, and that of Aq-C for instance, displayed in the right panel. Both halos
are abundant in stellar substructures of various forms that are the indications
of past accretion events. The disrupted substructures in these halos are not 
spherically distributed but lie along a particular direction, which is a very 
important property that has the power to potentially provide information 
regarding the origin of the progenitor satellites \citep{Helmi11}. While this 
has not been strictly quantified in the case of M31, looking at the left panel
of Figure~\ref{f:M31vsAq} it can easily be seen that most of the stellar debris 
is indeed aligned along the south-east to north-west direction. Similar 
anisotropy is readily observed in the simulations as well. Finally, and most 
importantly, many of the satellites that contribute to the build-up of the 
Aquarius stellar halos are massive enough to be likely parents of GCs, and their 
accretion can result in a formation of a GC system akin to that observed in M31 
as we discuss below.

\subsection{Creating mock globular cluster systems}
\label{ss:genGC}

We create mock GC systems of the final snapshot of the simulation around the 
main Aquarius halos using the dark matter particles  tagged with stellar 
properties by \citet{Cooper10}.  The resolution of these simulations does not 
allow for individual GCs to be resolved -- however, these objects would 
undoubtedly exist -- so we assume that certain tagged particles can also
represent the position and velocity of a GC contained within them. We are
unconcerned with the physical characteristics of the GCs (ages, metallicities,
luminosities, sizes), but are only interested in their spatial and kinematic
properties. By carefully selecting sets of these particles, we are effectively
building mock GC systems around the Aquarius galaxies.

\subsubsection{Galaxy scaling and placement of virtual observers}

Of all the Aquarius holos, the Aq-A halo has a total mass most similar to M31 
\citep{Fardal13}, and so we use it as a reference point. It also has the 
largest total mass and is the most radially extended in nature. In order
to explore the kinematic properties of GC systems in galaxies that have equal
total mass but different formation histories, we scale the positions and
velocities of the remaining Aquarius halos to match the mass of Aq-A. This is
done via the  scaling $M_{200}\propto r_{200}^3$ and
$M_{200}\propto v_{200}^3$ \citep{Helmi02}.

Prior to generating the mock GC systems, we ``observe'' each individual halo.
This means that we reformat the simulated data in way to emulate it being
obtained by a virtual observer placed 1000~kpc from the centre of the primary
galaxy. The 3D position of each star particle is transformed to a set of
pseudo-standard coordinates -- the particles are projected onto a plane passing
through the centre of mass of the primary halo and perpendicular to the line of
sight of the virtual observer -- while the 3D velocity is converted to a 
line-of-sight velocity according to the location of the observer. The right 
panel of Figure~\ref{f:M31vsAq} shows how Aq-C is seen from one such 
perspective.

We place 16 virtual observers around each Aquarius galaxy. Three are located
along the principal axis of the dark matter halo. The remaining 13 are
arranged in a regular pattern on a hemisphere centred on the target system. Of
these 13, one observer is placed at the arbitrarily chosen ``pole'' with polar
coordinates $(\theta,\phi) = (0,0)$ in the reference frame of that hemisphere.
The remaining 12 observers are grouped into three sets having polar
angles $\theta = (45, 90, 135)$~deg, and the 4 observers in each such set are
separated by 45~deg in azimuth, i.e. $\phi = (0, 45, 90, 135)$~deg.

\subsubsection{Build-up of a halo GC system}

Given the design of the Aquarius simulations, our mock GC systems are fully
populated by objects donated from the accreted dwarf galaxies, i.e. there is no
\emph{in situ} contribution. If such clusters exist, they are likely dominant
in the central regions of the galaxy, which we do not consider in this work. Of
course, it is possible that some of the \emph{in situ} clusters might be
``kicked'' to a larger radius by the complex interactions between the primary
galaxy and the accreting components, but such cases are expected to be rare.
Accreted satellites are eligible to donate GCs to the primary halo only
if they have a stellar mass higher than $10^6$~\Msun. The number of GCs donated
by each stream is related to its stellar mass.

In their model, \citet{Cooper10} chose to tag 1\% of the most bound
particles with stellar properties, successfully reproducing the
structural properties and the luminosity function of the surviving
satellites around the main halos. This choice, however, caused the
luminosity of Aq-A to be $\sim 100$ times fainter than that of M31
even though it has a similar total mass and its stellar halo is
approximately as extended as that of our neighbour. This effectively
implies a steeper $M_*/M_{\rm halo}$ relation at the faint end than the values 
used by e.g. \cite{Starkenburg13}, who also use the Aquarius suite. Had the
semi-analytic model been tuned in a way to make the halo of Aq-A have similar
luminosity to that of M31, the stellar streams around Aq-A would also be
brighter and of higher mass. This would in turn imply that they are  able
to carry more GCs than their current mass ranges suggest particularly compared
to the amount of GCs observed in dwarf galaxies
\citep[e.g.][]{Miller07,Peng08,Georgiev10, Veljanoski14}.

This is why we decided to populate mock GC systems according to the
following rules: If the predicted stellar mass of the progenitor according
to \cite{Cooper10} is between $10^6$ and $10^7$~\Msun, that progenitor will
contribute between 7 and 19 GCs, a value chosen uniformly at random. If the
accreted dwarf galaxy has a mass in the range of $10^7$-$10^8$~\Msun, it will
donate between 15 and 35 GCs, and if its mass is larger than $10^8$~\Msun\, the
contribution will range from 35 to 55 GCs. Taking this into consideration, the
number of GCs donated by each stream having a mass in the above ranges is
carefully scaled in such a manner that on average a mock GC system created
around Aq-A has as many members residing beyond 30~kpc in projection -- seen
from a randomly placed virtual observer around that galaxy -- as found in the
halo of M31. It is worth nothing that these scalings are consistent with
the recently measured correlations between the number and mass of globular
cluster systems and the host halo total mass by \cite{Harris15}.

As we want to create halo GC systems, only particles residing at projected
radii $\geq 30$~kpc may be chosen to represent GCs. To create a spatially
unbiased sample in the generation process, the number of GCs a stream can
donate to the galaxy, chosen based on its mass, is further multiplied by the
fraction of its particles that have projected radii $\geq 30$~kpc. This choice
is based on the assumption that GCs follow the light profile of their parent
dwarf galaxies or stellar streams. The limit of 30~kpc, chosen semi-arbitrarily,
marks the point where the stellar halo starts to dominate in the case of M31,
and is appropriate for our simulated galaxies given their mass and the radial
range to which the star particles extend to. An additional motivation comes from
observations of nearby galaxies, in which it is difficult to disentangle the
\emph{in situ} from the accreted stellar and GC populations within this radial
limit.

\subsubsection{Radial profile constraints, and GC selection algorithm}

\begin{figure}
\centering
\includegraphics[width=86mm]{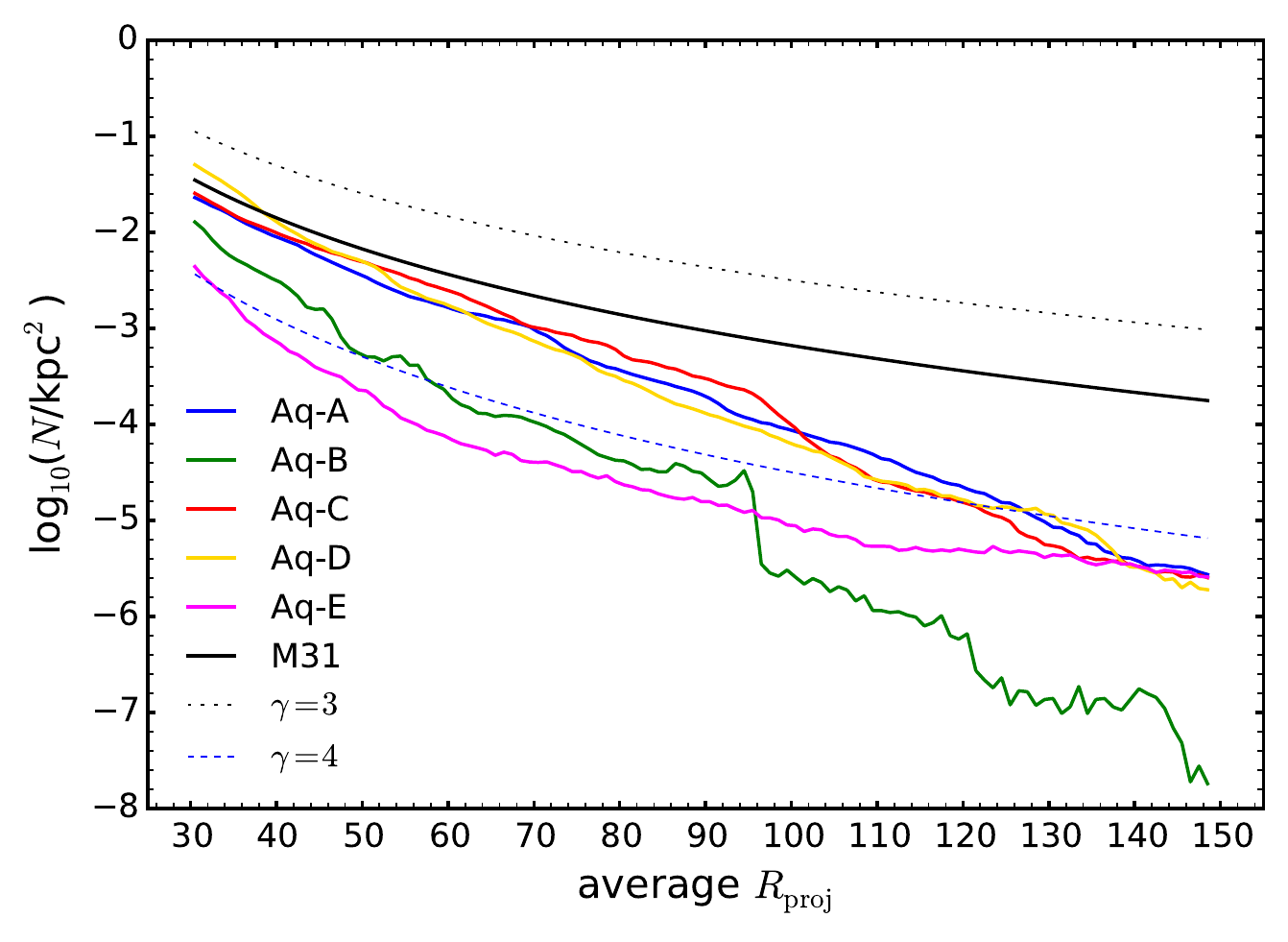}
\caption{Coloured lines represent the average stellar radial number density
profiles of each of the five Aquarius halos considered, as seen by 16 observers, 
3 of whom are located along the principal axis of the respective dark matter
halo, while the rest are arranged in a regular pattern spanning one hemisphere
centred on the target galaxy (see text for details). The black line shows the
stellar radial profile of metal-poor stars in the halo of M31, represented by a
power law with index of $-3.34$ \citep{Ibata14}. For reference, two generic
power laws having indices of $-3$ and $-4$ are plotted with dashed lines.}
\label{f:Aq-prof}
\end{figure}

Because we are trying to create GC systems analogous to that of M31, we require
the clusters in our simulations to follow the projected radial number density
profile of the star particles. In M31, the metal-poor halo stars and halo GCs
have similar radial number density profiles that have the same overall shape,
and features of interest are found at the same projected distances
\citep[see Fig. 11 in][]{Huxor11}. This is unsurprising given the likely common
origin of a dominant fraction of GCs and halo stars in this galaxy. Therefore,
we determine the projected radial number density profile for each of the primary
Aquarius galaxies self consistently. Figure~\ref{f:Aq-prof} displays the average
of 16 different profiles as seen by the virtual observers placed around each
simulated galaxy. For comparison, the figure also shows the radial number
density profile of metal-poor stars in the outer halo of M31, represented by an
intrinsic (not projected) power law with index of $-3.34$ \citep{Ibata14} and 
two generic power laws having indices of $-3$ and $-4$. From 
Figure~\ref{f:Aq-prof} it can easily be seen that the projected profiles of 
Aq-A, Aq-C, and Aq-D are similar to that of M31, while Aq-B and Aq-E are
considerably steeper and hence more compact in nature.

In summary, we generate a mock GC system following this algorithm: a
stellar halo is observed from the point of view of each
observer. Following this, the contribution of each stellar stream
towards the build-up of the GC system is decided. The stellar halo is
then divided into six equal projected radial bins between 30 and
150~kpc. The number of clusters in each radial bin is determined by
the stellar radial number density profile, as measured by the
respective observer. Stellar particles are then randomly chosen to
represent the positions and velocities of GCs from the appropriate
streams in a way to populate the radial bins following the conditions
detailed above. Throughout the selection process, no restriction is
put on how the GCs are selected in terms of their azimuthal
distribution. However, since the azimuthal distribution of the
accreted streams is not uniform, the GCs are also likely to depict
some amount of anisotropy.

\subsection{Kinematic modelling}
\label{ss:kinmod}

The main goal of this paper is to determine the kinematic properties of the
mock GC systems generated around the Aquarius galaxies. Following the example
set by \citet{Veljanoski14,Veljanoski15}, we construct a kinematic model and
then apply Bayesian tools to determine the most likely values for its free
parameters. Our model features two main components, one that constrains
the velocity dispersion of a GC system, while the other models the overall
rotation of the system, if present.

For the overall rotation component, we adopt the model described in detail in
\citet{Cote01}, which has the form:

\begin{equation}
v_{\rm p}(\theta) = v_{\rm sys} + A\,{\rm sin}(\theta - \theta_{0})
\label{e:rot}
\end{equation}
where $v_{\rm p}$ and $\theta$ are the radial velocities and
position angles of the GCs, and $\theta_0$ the position angle of the rotation
axis of the GC system respectively. The rotational amplitude is denoted $A$,
while $v_{\rm sys}$ represents the systemic radial motion of the GC system as a
whole. For each halo, $v_{\rm sys}$ is the mean velocity of all particles
located within 30~kpc of the centre, as seen from each observer individually.
This is equivalent to adopting the systemic motion of the host galaxy as the
mean velocity of a GC system and is a common practice when determining the
kinematic properties of real GC systems.

We assume that the velocity dispersion can be well represented by a Gaussian
which does not vary as a function of projected distance from the centre of the
galaxy. Mathematically, it can be expressed as:
\begin{equation}
\sigma^2  = (\Delta v)^2 + \sigma_0^2
\label{e:constdisp}
\end{equation}
where $\sigma_0$ is the intrinsic velocity dispersion of the GC system around a
particular galaxy and $\Delta v$ are the uncertainties in the radial velocities
of the GCs. In this work, we do not aim to simulate the observations and the
data acquisition process; assign observation uncertainties to the
radial velocities of the GC particles to be $\pm 15$~\kms, a typical value
expected when observing a cluster near the peak of the GC luminosity
function observed with a 4-meter ground-based facility, with a signal-to-noise
ratio of $3 < S/N < 8$.

Putting together equations (\ref{e:rot}) and (\ref{e:constdisp}), we are able to
construct our kinematic model which has the form:
\begin{equation}
p_{i,{\cal M}}(v_i, \Delta v_i|v_{\rm p}, \sigma) = \frac{1}{\sqrt{2\pi\sigma^2}}  \:\: \exp{\left(-\frac{(v_i - v_{\rm p})^2}{2\sigma^2}\right)}
\label{q:kinmod}
\end{equation}
where, as described earlier, $v_{\rm p}$ and $\sigma$ are the overall rotation
and velocity dispersion of the GC system, and $v_i$ are the radial velocities
of the GCs. The likelihood function for the above model is therefore:
\begin{equation}
p_{{\cal M}}(D|\Theta) = {\cal L}_{{\cal M}}\left(v, \Delta v, \theta | A,\theta_0,\sigma_0 \right)  = \prod_i p_{i,{\cal M}} 
\label{e:likelihood}
\end{equation}
in which $v$, $\Delta v$ and $\theta$ are the observed properties of the
GCs, while $A$, $\theta_0$, $\sigma_0$ are the free parameters of the model. The
index $i$ loops over each GC. In the following analysis, we assume flat priors
for all free parameters. One advantage of our approach is that we characterize
the rotation (if present) and dispersion simultaneously, and thus avoid any 
biases that could be introduced when trying to model these properties in 
succession.

The likelihood function is computed via a brute-force method over
a regular grid shaped by the free parameters of the kinematic model. In these
computations, the rotation amplitude $A$ and the velocity dispersion $\sigma_0$
span the ranges from 0 to 150~\kms and from 0 to 200~\kms respectively, both
with resolution of 3~\kms. The rotation axis $\theta_0$ is searched in the
interval between 0 and $2\pi$~rad with a step size of 0.15~rad. We carefully
tested the choice of parameter ranges and the sampling, and find them to give
excellent balance between computational speed and resolution of the likelihood
function and the posterior probability distributions. Once the likelihood
function is determined, we calculate the marginalized posterior probability
distribution function for each free parameter.

\section{Kinematic analysis}

The purpose of this experiment is to study the kinematic properties of mock GC
systems as if they were observed in the same fashion as the GC system of M31.
In order to get a statistically more robust estimate, and to reduce Poisson
noise effects, we independently generate 100 mock GC systems for each of the 16
observers placed around the Aquarius halos via our prescription detailed in
Section~\ref{ss:genGC}. The multiple observers ensure that we can analyse the
halo from many different perspectives, since in reality galaxies have many
different orientations that influence the measurements. Having observed the
Aquarius galaxies from different perspectives allows us to compare the kinematic
properties of the associated mock GC systems as a function of galaxy
orientation. Iterating the procedure 100 times per observer helps to determine
the representative properties of a typical GC system that could exist in that
particular halo.

We apply our Bayesian methodology as described in Section~\ref{ss:kinmod} to
each of these systems, and calculate the most likely values for the amplitude 
and axis of rotation and for the velocity dispersion. The degree of rotation in 
each GC system is typically classified by the ratio between the maximum rotation 
amplitude and the velocity dispersion $\left( A/\sigma_0 \right)$, also called 
the kinematic ratio. If $\left( A/\sigma_0 \right) < 0.5$, the system is deemed 
to be non-rotating (or slowly rotating); 
if $ 0.5<\left( A/\sigma_0 \right)<1.0$,
the system has a non-negligible rotation signature; 
if $\left(A/\sigma_0\right)>1.0,$ the system is a fast rotator. For reference,
the kinematic ratio of the halo GC system of M31 is $0.67$ \citep{Veljanoski14}.
In that study, the authors applied a similar but not identical kinematic model
since they also considered how the velocity dispersion varies as a function of
projected radius. Applying the exact model used in this study to the data
presented in \citet{Veljanoski14} we find the kinematic ratio of the M31
outer halo GC system to be $0.93$.

\begin{figure*}
\centering
\includegraphics[scale = 0.75, page = 1]{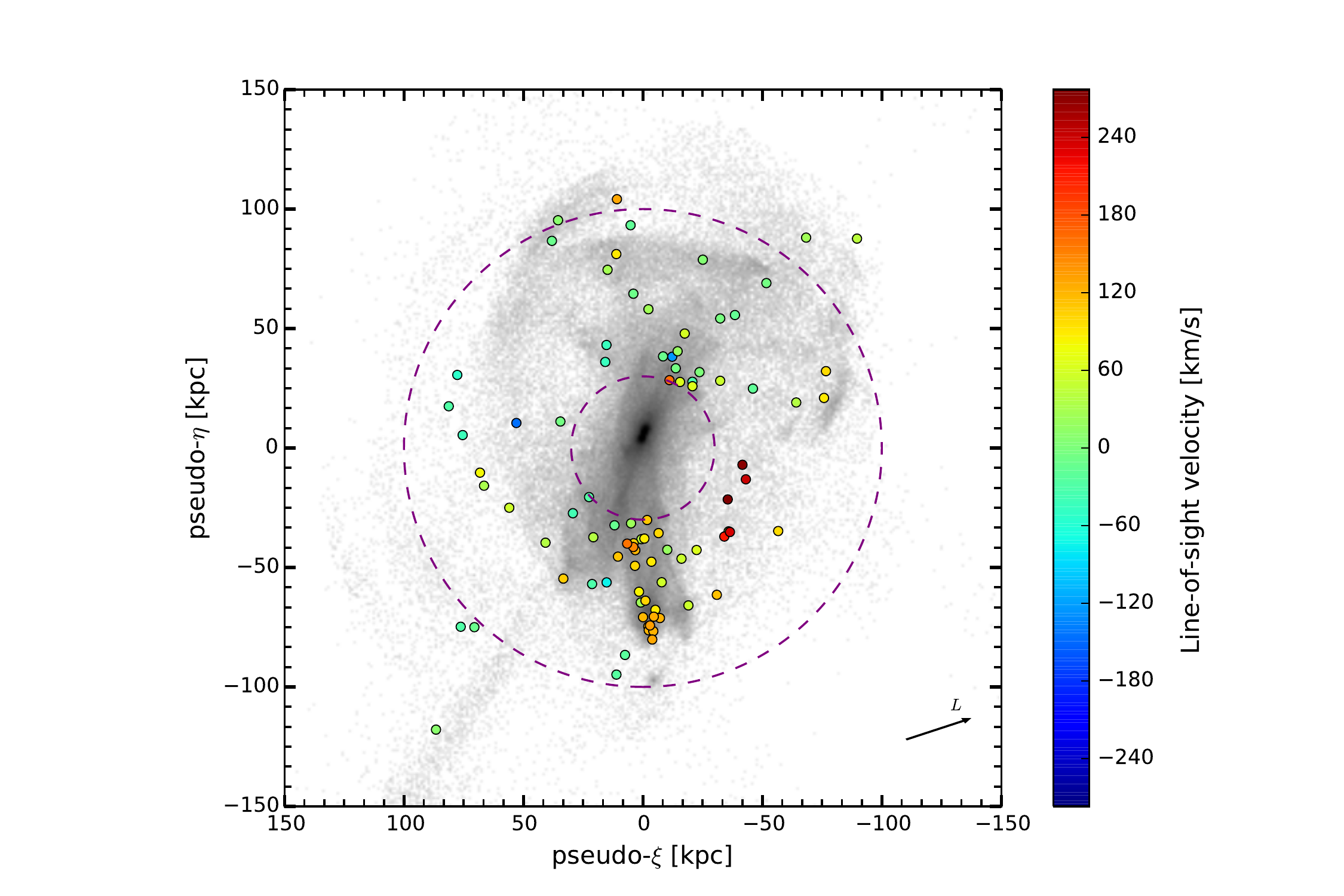} \\
\includegraphics[scale = 0.50, page = 2]{fig3.pdf}
\caption{Top panel: Highly rotating
  $(A/\sigma_0 = 1.38)$ mock GC system around Aq-A, observed from
  perspective $(\theta,\phi)=(0,0)$. The colours of the points marking
  the positions of the GCs correspond to their radial velocity as seen
  by the virtual observer. The arrow in the bottom right corner of
  the map indicates the orientation of the angular momentum of the
  stellar halo as seen on the plane of the sky. Four diagnostic plots
  are displayed below each of the two maps. The top left 
  shows the line-of-sight velocities of the GCs as a function of their
  position angle. The best fit rotation curve is fitted through the
  plots according to Equation~(\ref{e:rot}). The horizontal dashed
  line represents the systemic velocity of the halo. The remaining
  three plots (moving clockwise) display the posterior probability
  distribution functions for the amplitude, velocity dispersion, and
  rotation axis of the mock GC system as determined by our Bayesian
  machinery. The peak of each distribution and the $1-\sigma$
  uncertainties are indicated by the dashed and solid vertical
  lines, respectively.}
\label{f:exampleMap_r}
\end{figure*}

\begin{figure*}
\centering
\includegraphics[scale = 0.75, page = 1]{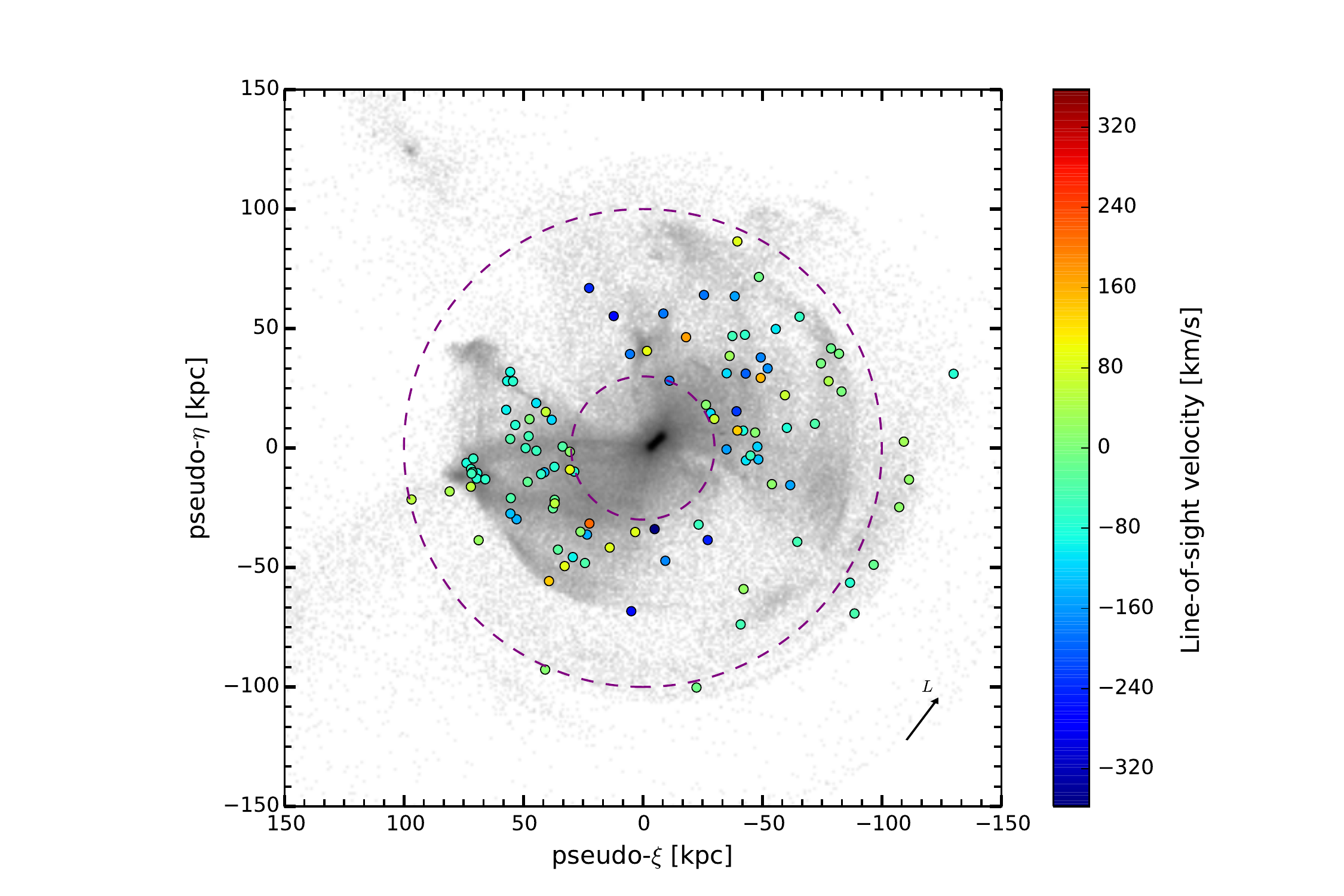}  \\
\includegraphics[scale = 0.50, page = 2]{fig4.pdf}
\caption{Same as Figure~\ref{f:exampleMap_r}, but for a mock GC system observed
from $(\theta,\phi)=(135,0)$, which shows no rotation signal
$(A/\sigma_0 \approx 0)$.}
\label{f:exampleMap_nr}
\end{figure*}

To illustrate how our cluster selection and kinematic analysis algorithms work
in practice, we show two examples in Figure~\ref{f:exampleMap_r} and
\ref{f:exampleMap_nr}. Fig.~\ref{f:exampleMap_r} shows a mock GC system around
Aq-A observed from the perspective $(\theta,\phi)=(0,0)$, which exhibits a
notable rotation signal $(A/\sigma_0 = 1.38)$. The system shown in 
Figure~\ref{f:exampleMap_nr}, observed from the perspective 
$(\theta,\phi)=(135,0)$, does not appear to be rotating 
$(A/\sigma_0 \approx 0)$. In addition, four diagnostic plots are shown
below each of the example maps. Three of these give the posterior
probability distribution functions for the free parameters of our kinematic
model, while one is a plot of the line-of-sight velocities of the GCs versus
their position angles as seen on the sky. On these plots one can clearly see
the presence, or absence, of a rotation signature.

In order to check whether the azimuthal distribution of the GCs around each 
Aquarius halo has any effect on the kinematic results we derive in our analysis, 
we select a subsample of our mock systems that have similar azimuthal 
distributions to the globulars around  our neighbour. This is done in the 
following manner: we split the halo of M31 into four equal quadrants, the 
borders of which are the natural major and minor axis of its disk, as seen on
the sky. Counting the number of GCs in each region, we find that one quadrant 
has $\sim1\sigma$ more and one quadrant has $\sim1\sigma$ less than the average 
number of GCs contained in all four quadrants. We look for the same spatial 
property in our mock systems by separating the halos into four quadrants where 
the borders are marked by the lines due north-south and east-west, as seen on 
the arbitrarily oriented sky by each observer. We then select those systems 
which feature two quadrants that deviate from the mean by $1\sigma$: one 
quadrant is required to be underpopulated and the other overpopulated with 
clusters. On average, nearly half of all generated GC systems around the five 
Aquarius halos satisfy this criteria. Although the orientation of the quadrants 
is chosen arbitrarily, we note that this particular choice has no effect on the 
results. Choosing a different, random orientation for the quadrants does not 
change the results that we discuss below.

Table~\ref{t:kinres} presents the main bulk kinematic properties of the five
Aquarius galaxies that we consider in this work. The table shows that on average
little rotation signal is detected when  attempting to constrain the kinematics
of a GC system by viewing it from a random direction. However, when looking from
the ``correct'' point of view, one can indeed detect a significant degree of
rotation. In what follows we describe the results for each halo individually. We
focus on the bulk kinematic properties of the GC systems generated around each
halo, while also investigating how well the true kinematic properties of the GC
systems are recovered by our analysis, and to what degree we can constrain the
direction of of the angular momentum vector of the stellar halos by studying
the motions of the associated GCs.

\begin{table*}
\centering
\caption{ Bulk kinematic properties of the mock GC systems in the Aquarius
galaxies. The column titled $\langle N_{\rm GC} \rangle $ gives the average
number of GCs comprising a mock system around an Aquarius halo. The columns
labelled  $A/\sigma_0$ and $p(A/\sigma_0 \geqslant 0.93)$ show the the average
kinematic ratio regardless of the observer's position and the probability of
observing a kinematic ratio at least as large as that of the M31 halo GC
population, respectively. The values quoted are nearly identical to those
obtained when taking into account only those GC systems that have  similar
azimuthal distributions to the halo GCs around M31 (see text for details). The
columns with the subscipt ``max''  show statistics only for the perspective from
which the largest rotation signal is detected.}
\label{t:kinres}
\begin{tabular}{lccccc}
 \hline
 \hline
 Halo & $\langle N_{\rm GC} \rangle$ & $A/\sigma_0$ & $p(A/\sigma_0 \geqslant 0.93)$ & $(A/\sigma_0)_{\rm max}$ & $p_{\rm max}(A/\sigma_0 \geqslant 0.93)$\\
 \hline
 Aq-A & 84  & 0.3    & $<0.1$ & 0.8 & 0.3 \\
 Aq-B & 30  & $<0.1$ & 0.1    & 0.8 & 0.3  \\
 Aq-C & 106 & 0.4    & 0.1    & 1.0 & 0.6 \\
 Aq-D & 94  & 0.2    & $<0.1$ & 0.5 & 0.1 \\
 Aq-E & 20  & 0.7    & 0.4    & 1.9 & 0.9  \\
 \hline
\end{tabular}
\end{table*}

\subsection{Aq-A}
\label{ss:A}

Aq-A has a virial mass of $1.84 \times 10^{12}$~\Msun, the highest of the
Aquarius halos. Out of the seven streams that have stellar masses higher than
$10^6$~\Msun\ and abide in the outer regions of this halo, six donate at least 
10 GCs, while two of them contribute more than 17 GCs on average. A typical mock
GC system of Aq-A comprises 84 members. The mean uncertainty with which the
rotation amplitude, rotation axis, and velocity dispersion are determined is
17~\kms, 42~deg, and 7~\kms\, respectively\footnote{When the rotation present in
the system is close to null, the determination (and corresponding uncertainty)
of the orientation of the rotation axis is naturally large.}.

\begin{figure}
\centering
\includegraphics[width=86mm]{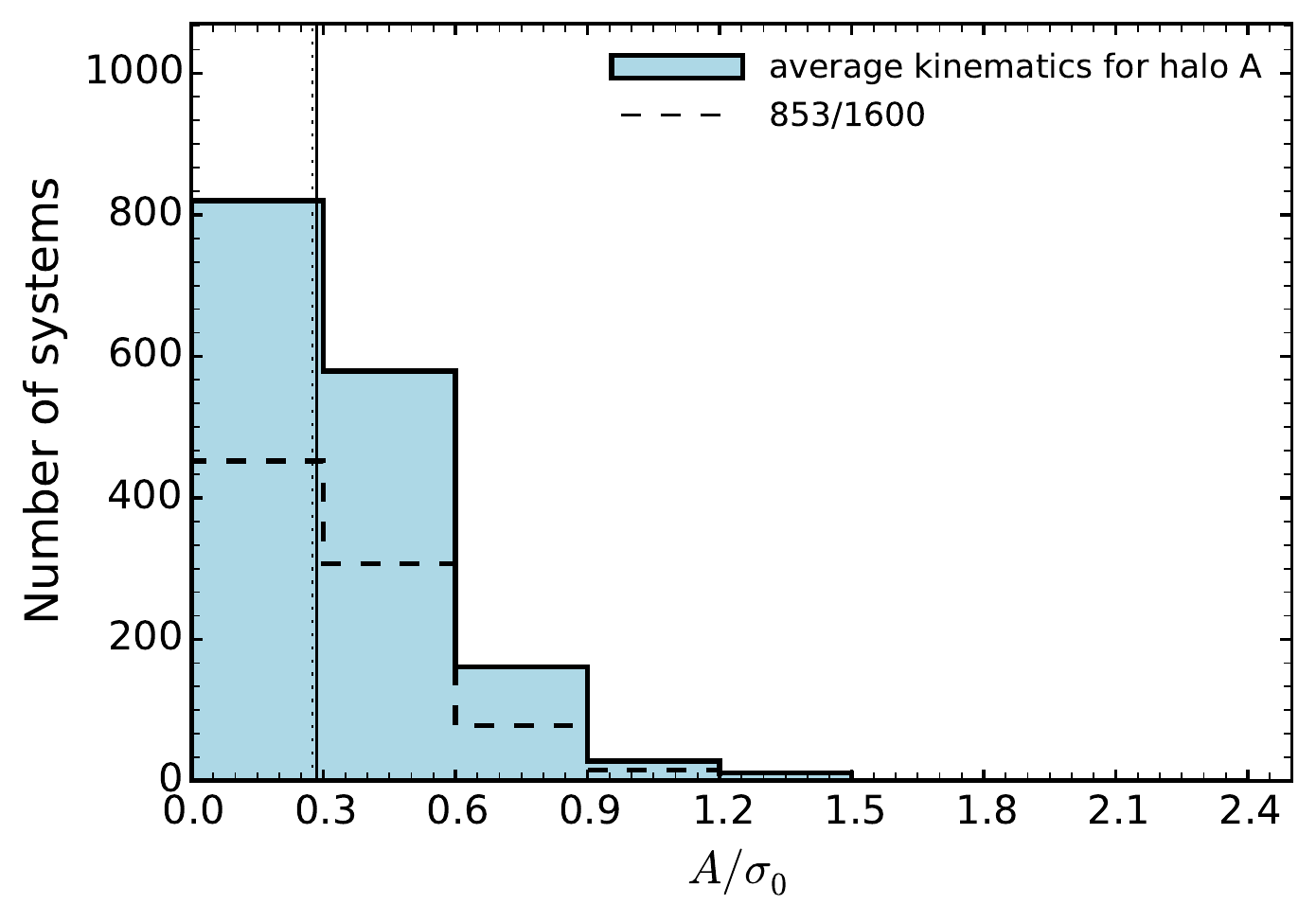}
\caption{Kinematic ratio $A/\sigma_0$ for all generated mock GC systems in the 
halo of Aq-A, regardless of viewing angle. The dashed histogram shows the 
subsample of systems that have similar azimuthal distributions to the halo GC 
system of M31 (see text for details). The solid and dashed vertical lines 
represent the mean kinematic ratio measured for all systems and for the 
azimuthally selected subsample of systems, respectively.}
\label{f:avghistA}
\end{figure}

Figure~\ref{f:avghistA} shows a histogram of the kinematic ratio for all mock GC
systems generated in the halo of Aq-A, regardless of where the virtual observers
are placed. It is clear that the mean kinematic ratio for all mock GC systems is
only 0.3, as shown by the vertical solid line in the figure. In fact, the chance
of observing a mock halo GC system that has a kinematic ratio as large as that
observed in M31 or larger when Aq-A is observed from a random perspective is, on
average, lower than 0.1\%.

\begin{figure*}
\centering
\includegraphics[scale = 0.6]{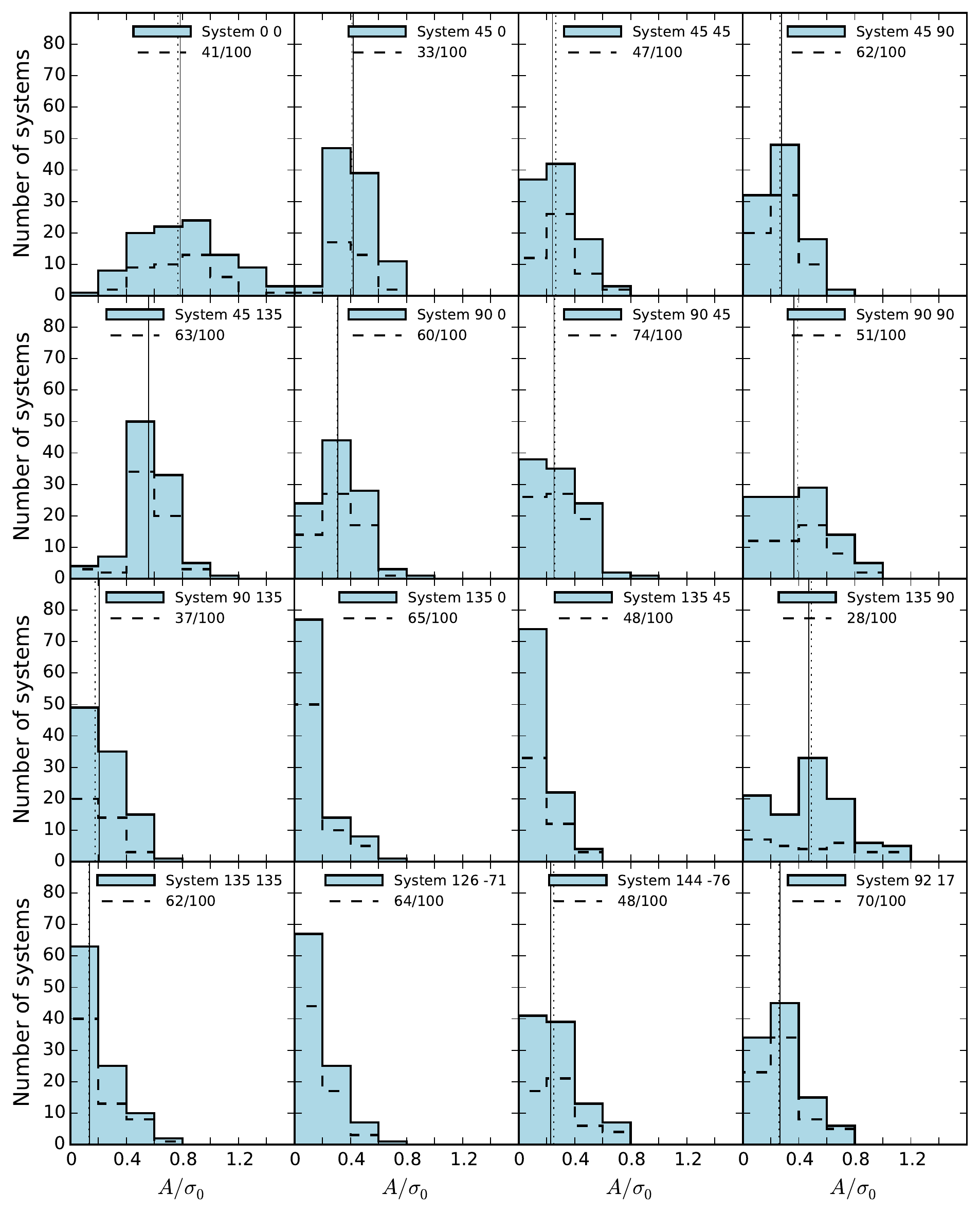}
\caption{Kinematic ratio of the generated GC systems around Aq-A for each
of the 16 individual observers in histogram format. The dashed histograms show
the subsample of systems that have similar azimuthal distributions to the
GCs in the halo of M31. The solid and dashed vertical lines represent the mean
kinematic ratio measured for all systems and for the azimuthally selected 
subsample of systems, respectively.}
\label{f:histfigA}
\end{figure*}

This kinematic picture changes significantly once we look at each observer
individually. Figure~\ref{f:histfigA} displays in histogram format the kinematic 
ratio of the mock GC systems as measured by the 16 observers placed around Aq-A. 
The histogram clearly shows that the results are highly dependant on the 
location of the observer. For instance, the observer located at 
$(\theta,\phi) = (0,0)$ in an arbitrary oriented reference frame centred on
Aq-A, detects the highest rotation signal of all the observers, while the
observer located at $(\theta,\phi) = (135,0)$ measures no rotation signature.
From the favourable perspective, $(\theta,\phi) = (0,0)$, the probability of
observing a GC system that has  a $A/\sigma$ at least as high as the halo GC
system around M31 is 30\%. We note that when determining the kinematic 
parameters for the GC systems observed from this perspective, the corresponding
uncertainties for the rotation amplitude, rotation axis, and velocity
dispersion are 16~\kms, 15~deg, and 6~\kms,\ respectively.

The histograms drawn with dashed lines in
Figure~\ref{f:avghistA}~and~\ref{f:histfigA} correspond to the subsamples of
GC systems that have azimuthal distribution similar to the halo GC system of
M31. The figures show that this subsample of systems does not deviate from the 
trends set by the overall population. The same holds true for the other four 
halos we are exploring.

\begin{figure}
\centering
\includegraphics[scale = 0.6]{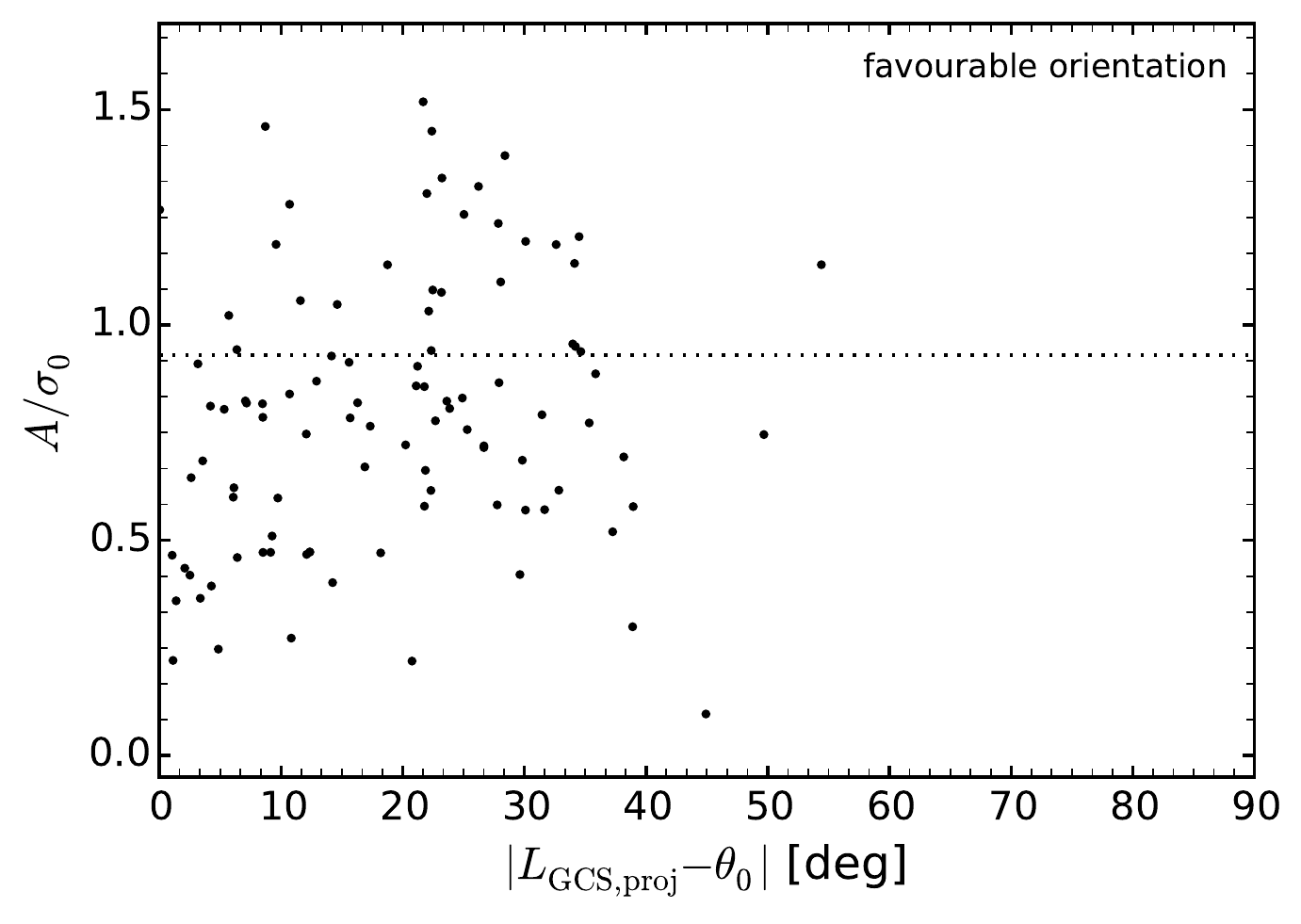} \\
\includegraphics[scale = 0.6]{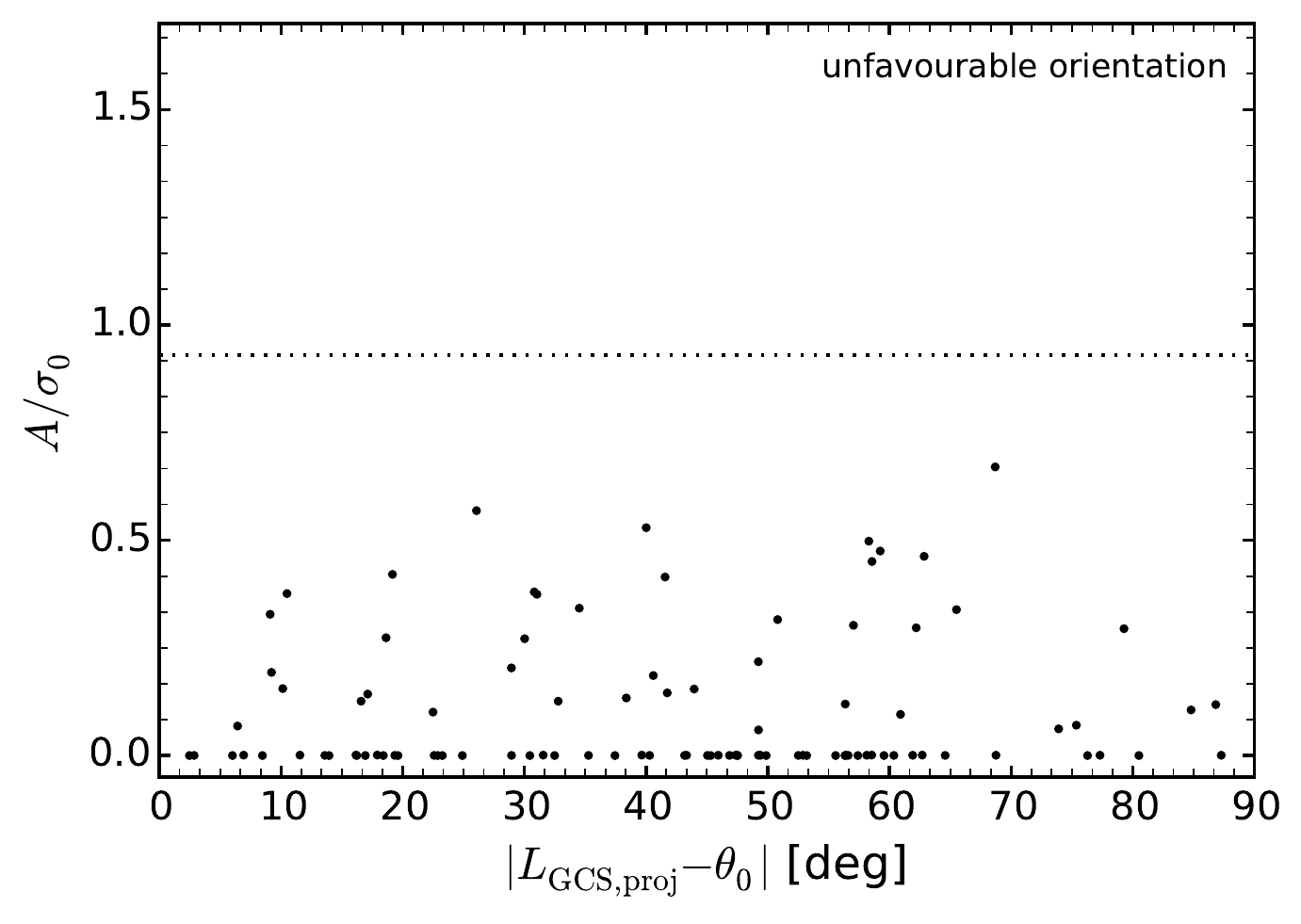} \\
\includegraphics[scale = 0.6]{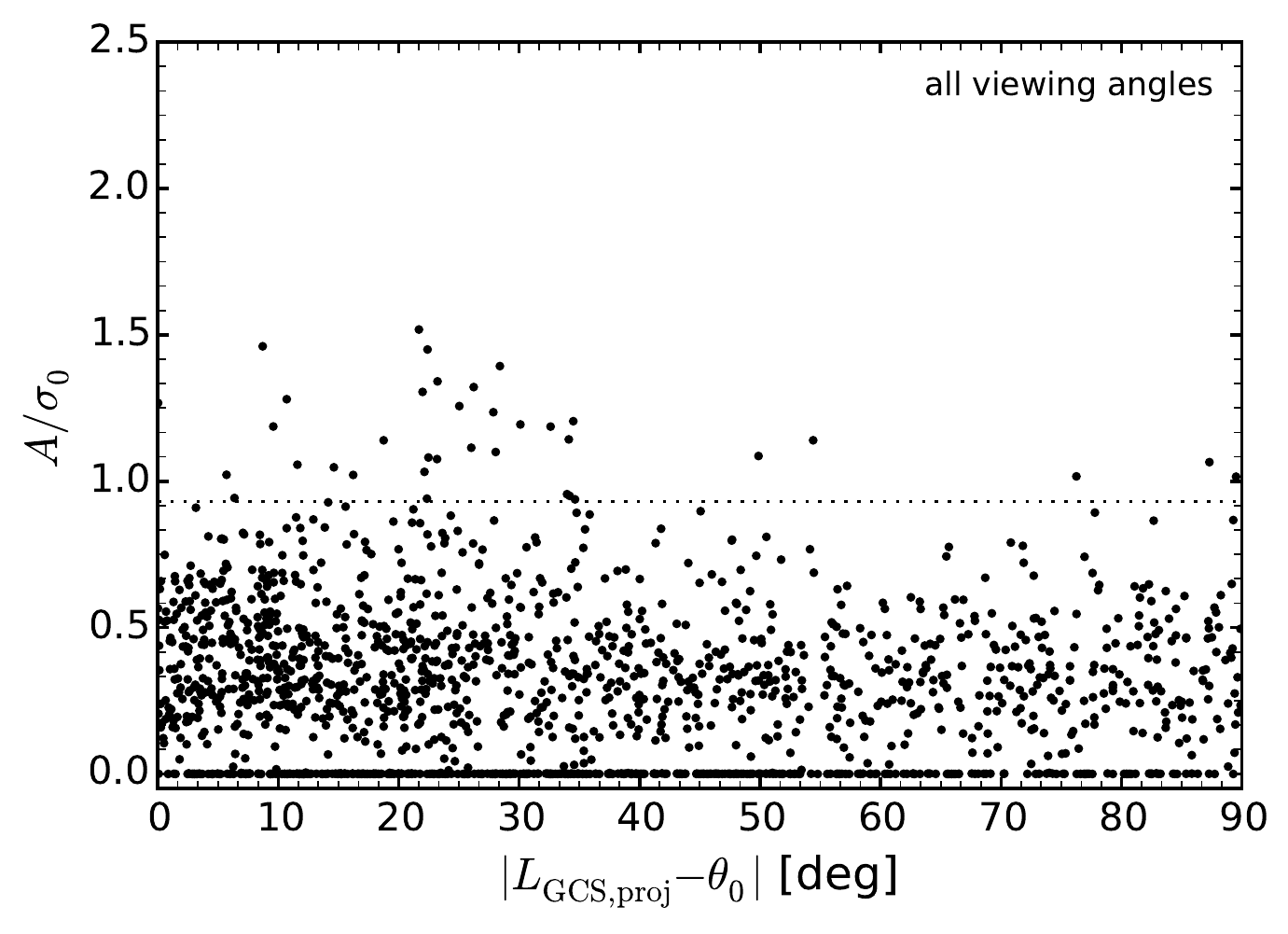}
\caption{Kinematic ratio of the mock GC systems in the outer halo of Aq-A
versus the difference between the position angle of their angular momentum
projected on the plane of the sky and the position angle of the rotation axis as
determined by our kinematic analysis. The top panel shows the systems seen from
an observer located at $(\theta,\phi)=(0,0)$  who detects the maximum rotation
signal, and the middle panel shows the systems seen by an observer located at
$(\theta,\phi)=(135,0)$ who detects no such structured motion. The bottom
panel displays all systems regardless of the viewing perspective. We note that 
the projected angular momentum and rotation axis agree well for systems that 
exhibit a high degree of rotation.}
\label{f:angmomA}
\end{figure}

In Figure~\ref{f:angmomA} we  investigate how well the true kinematic properties 
of the mock GC systems are recovered by our kinematic analysis. Here we plot the 
kinematic ratio of the mock GC systems around Aq-A versus the difference between 
the position angle of the true angular momentum of each system projected onto 
the plane of the sky, and the position angle of the axis of rotation as 
determined by our kinematic analysis. The top panel in this figure shows the 
systems observed from the perspective from which the detected rotation signal is 
highest $(\theta,\phi)=(0,0)$, while the systems displayed in the middle panel 
are observed from a perspective with no such signal, $(\theta,\phi)=(135,0)$. 
The bottom panel in this plot shows all mock GC systems, irrespective of the 
viewing perspective. The dotted horizontal line in these panels represents the 
kinematic ratio of M31's outer halo GC system. For the cases in which the 
determined kinematic ratio is at least as high as that of the outer halo GC 
system of M31, the difference between the projected angular momentum and the 
measured rotation axis is less than 35~deg for 86\% of the systems, i.e. for 
35 of the 40 systems that have $A/\sigma_0 \geq 0.93$. Naturally, when the 
rotation signal is zero or spurious,  as it is in the middle panel in 
Figure~\ref{f:angmomA}, such a comparison cannot be made.

\begin{figure}
\centering
\includegraphics[scale = 0.6]{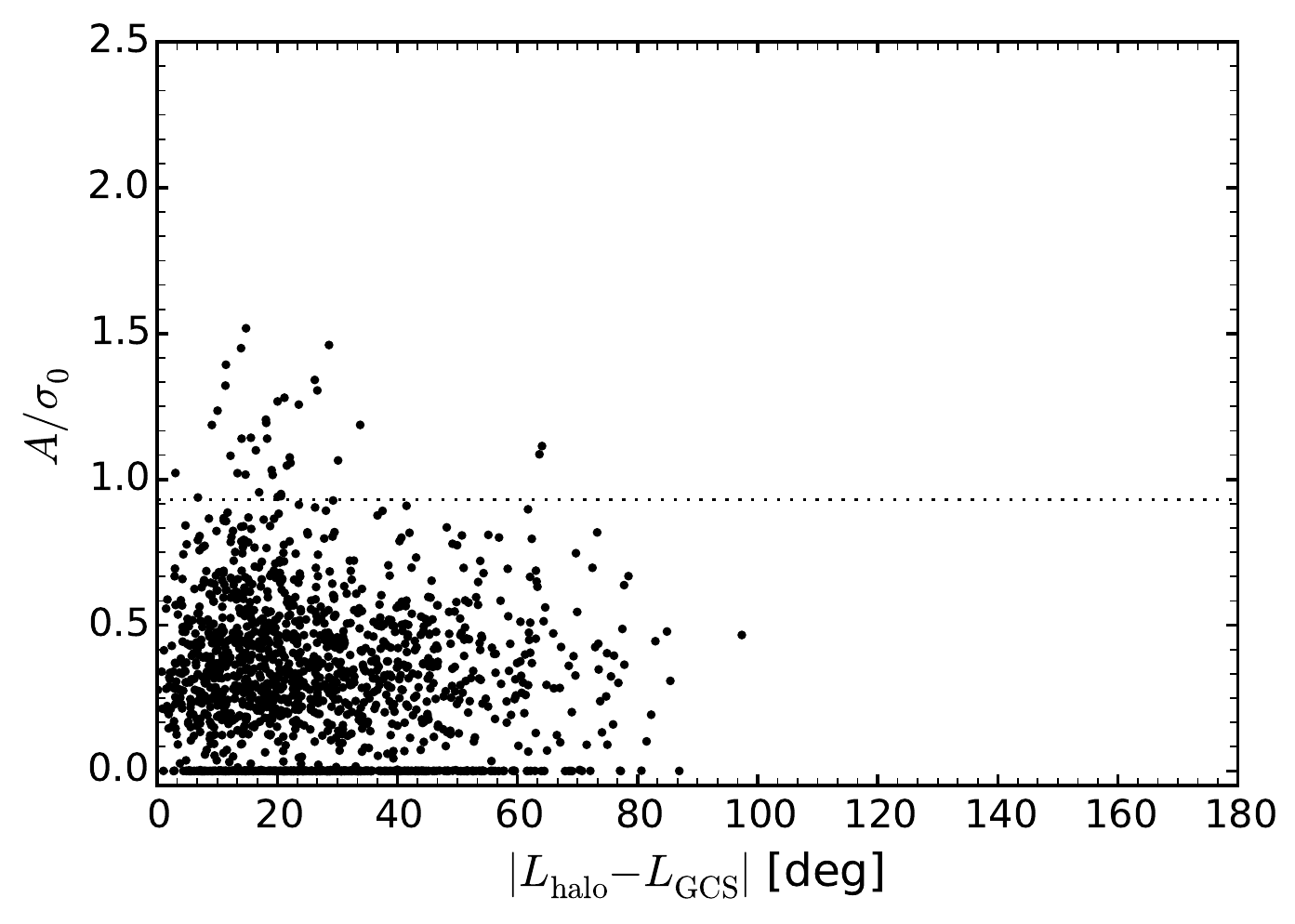}
\caption{Kinematic ratio of the mock GC systems around Aq-A versus the
difference in angle between their true 3D  angular momentum vector
and the 3D angular momentum vector of the outer halo star particles,
regardless of the viewing perspective. The misalignment between the GC and halo
angular momentum is relatively small, especially for high kinematic ratios.}
\label{f:angmom3DA}
\end{figure}

The idea that a GC system can be used to constrain the kinematics of the halo it 
is inhabiting is further supported by Figure~\ref{f:angmom3DA}. Here we plot the 
kinematic ratio of all our mock GC systems, regardless of the viewing 
perspective, versus the angular difference between the true 3D angular momentum
vector of those GC systems and the net angular momentum of all star particles 
located beyond 30~kpc,  i.e. in the outer halo. We find that for 86\% of the 
cases, the mock GC systems constrain the orientation of the angular momentum of 
the stellar halo within 65~deg of the true direction. This number is reduced to 
only 25~deg if we consider GC systems having $A/\sigma_0 \geq 0.93$.

\subsection{Aq-B}
\label{ss:B}

In contrast to Aq-A, Aq-B is the least massive Aquarius halo; it has a virial
mass of $0.82 \times 10^{12}$~\Msun\ prior to scaling. It is also the most
compact of the halos we are considering. As a result, there are only four 
streams that can donate GCs to the halo of this galaxy. Two of these streams 
contribute $\sim 10$ while the other two supply $\sim 6$ members to the
mock cluster systems. The ensembles have 30 members on average. The mean
uncertainty when determining the rotation amplitude,  axis, and velocity
dispersion are 32~\kms, 50~\kms, and 12~\kms, respectively.

\begin{figure*}
\begin{tabular}{@{}ccc}
\includegraphics[scale = 0.41]{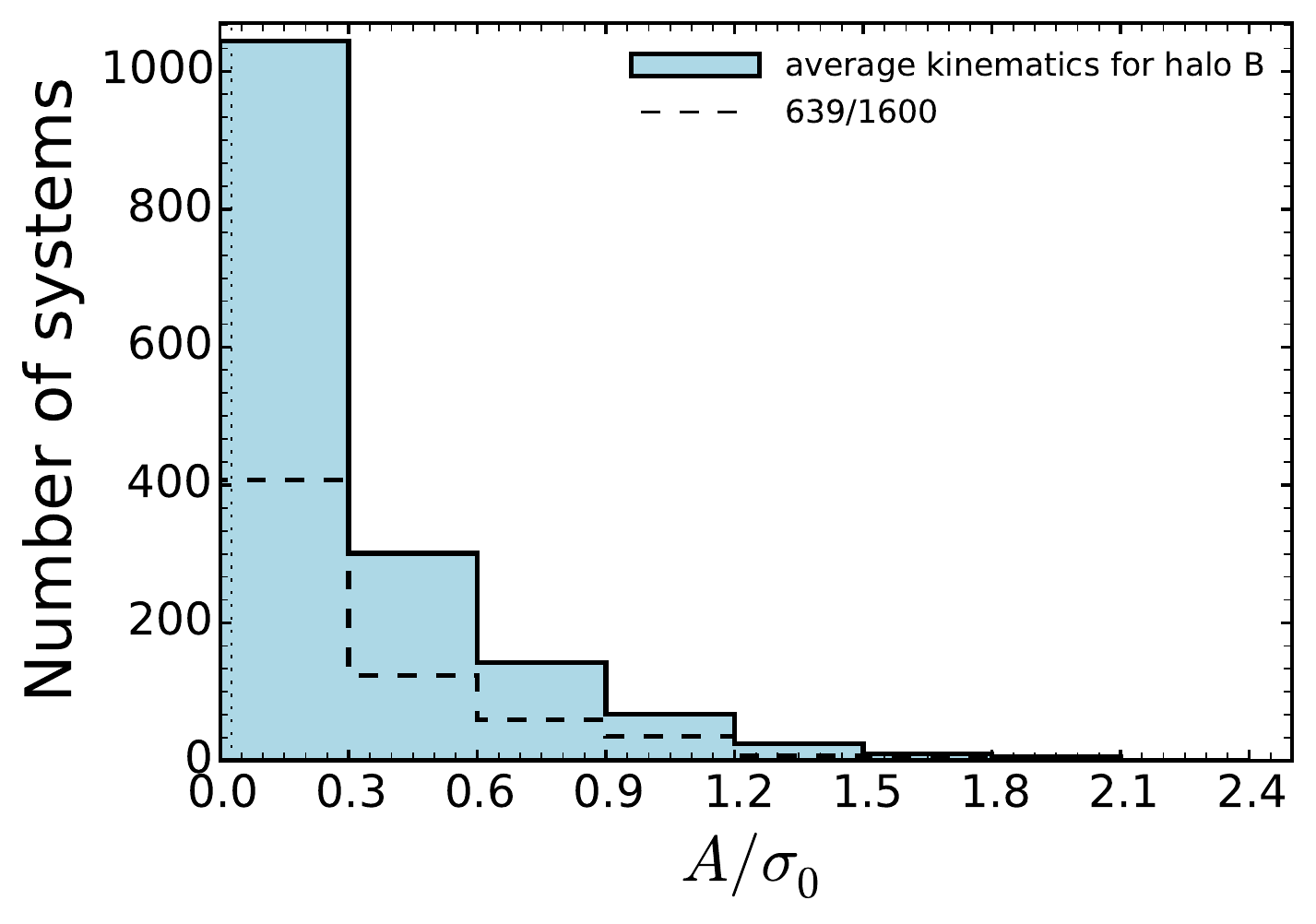} &
\includegraphics[scale = 0.41]{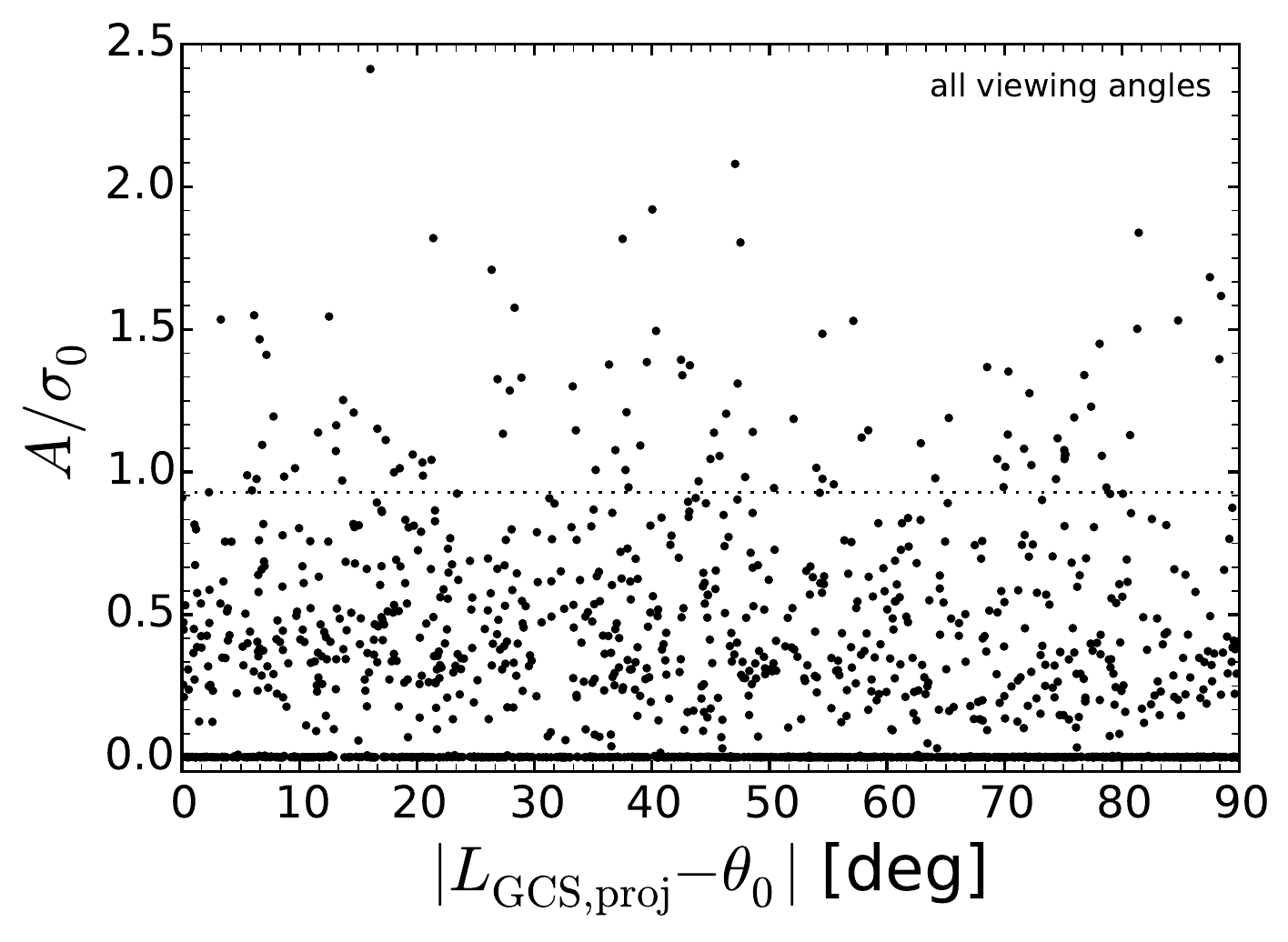} &
\includegraphics[scale = 0.41]{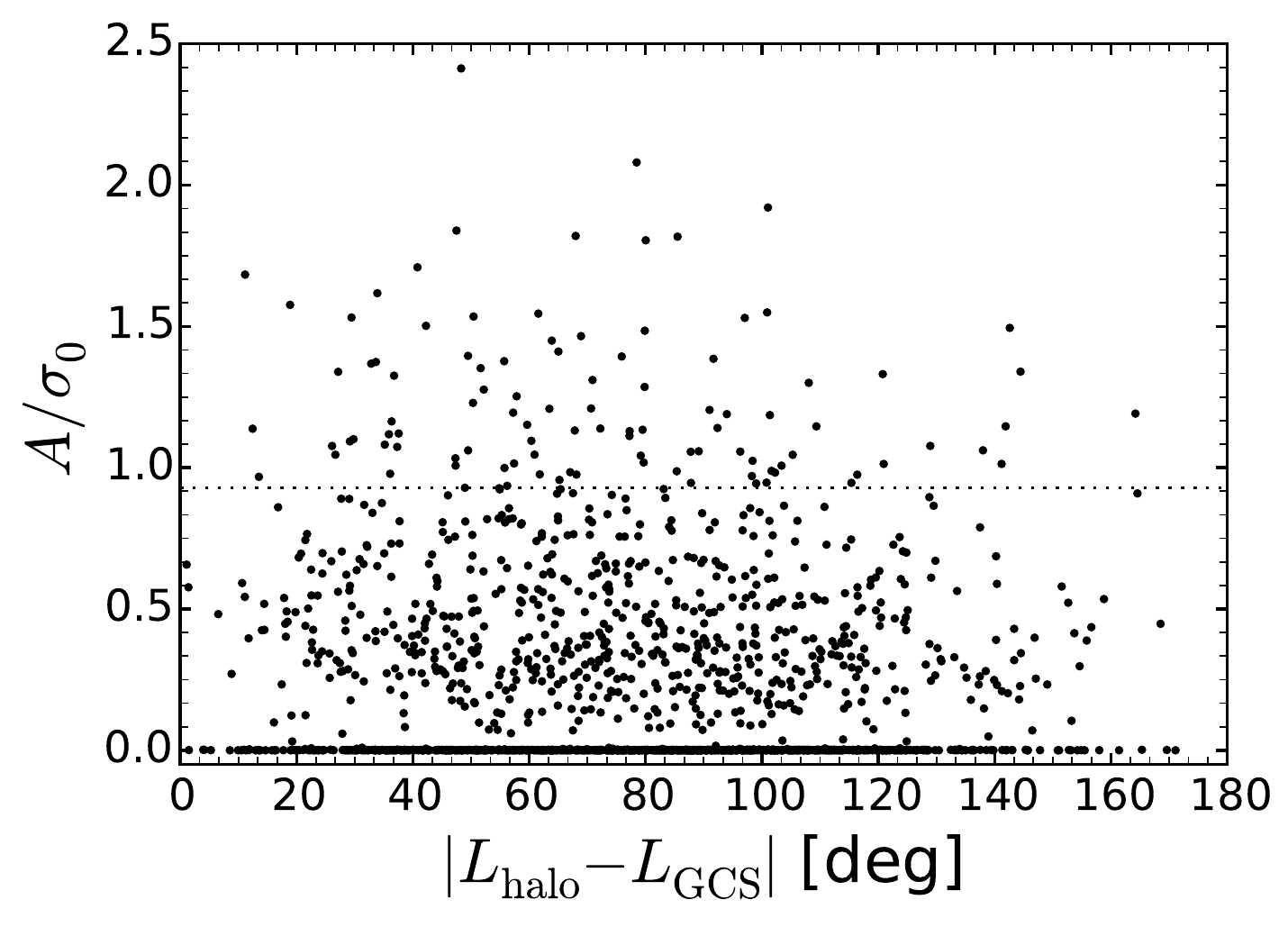} \\
\includegraphics[scale = 0.41]{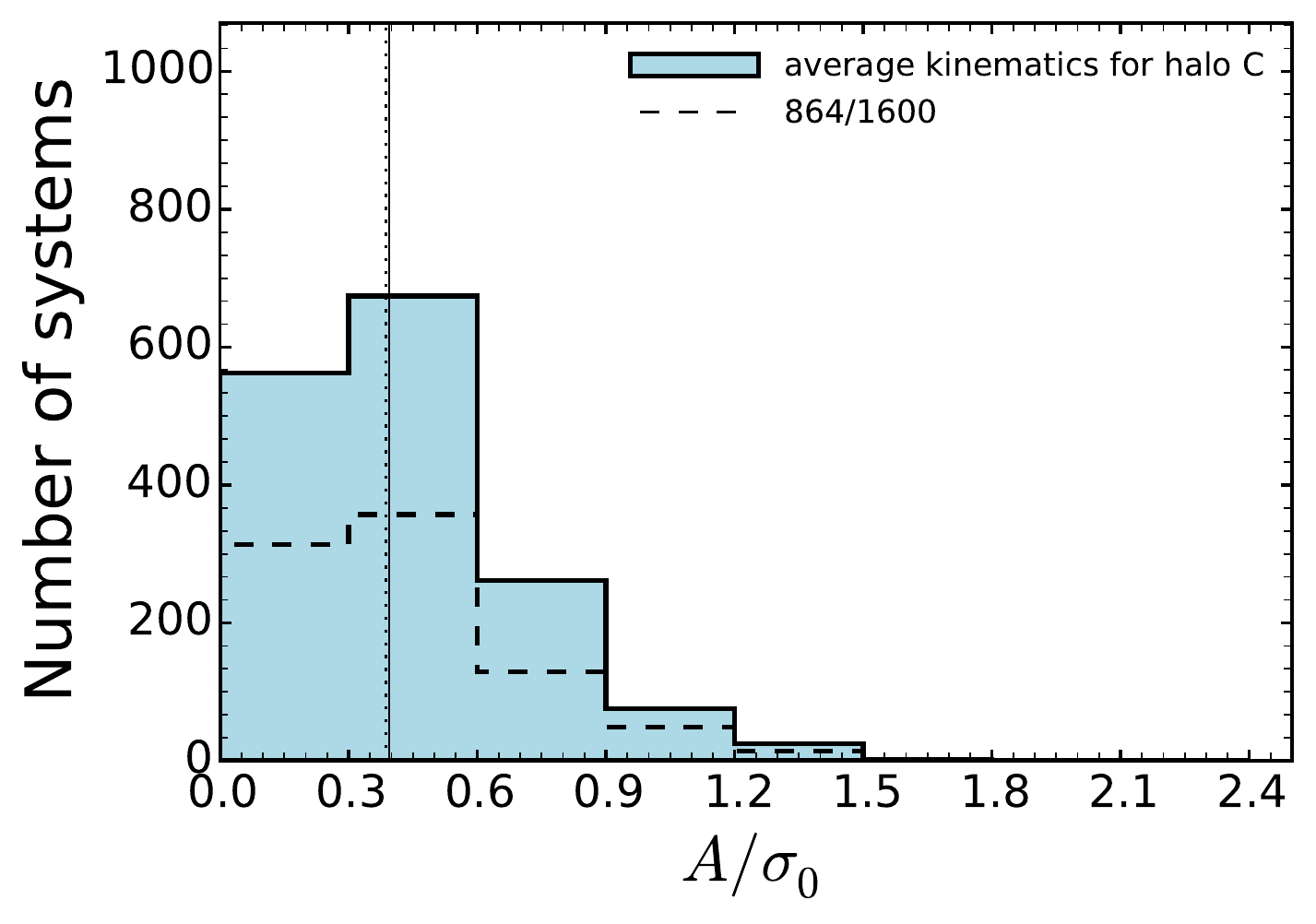} &
\includegraphics[scale = 0.41]{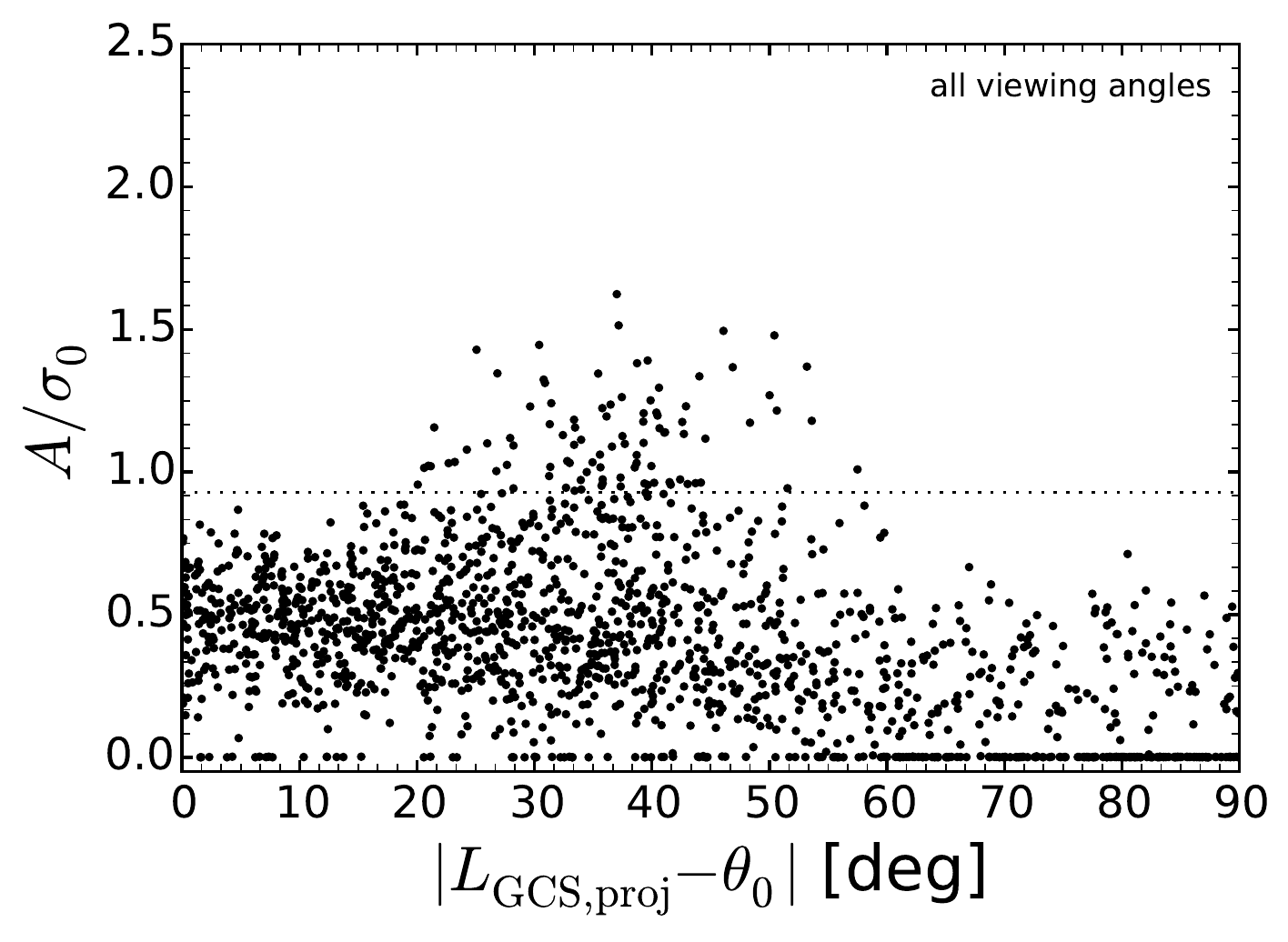} &
\includegraphics[scale = 0.41]{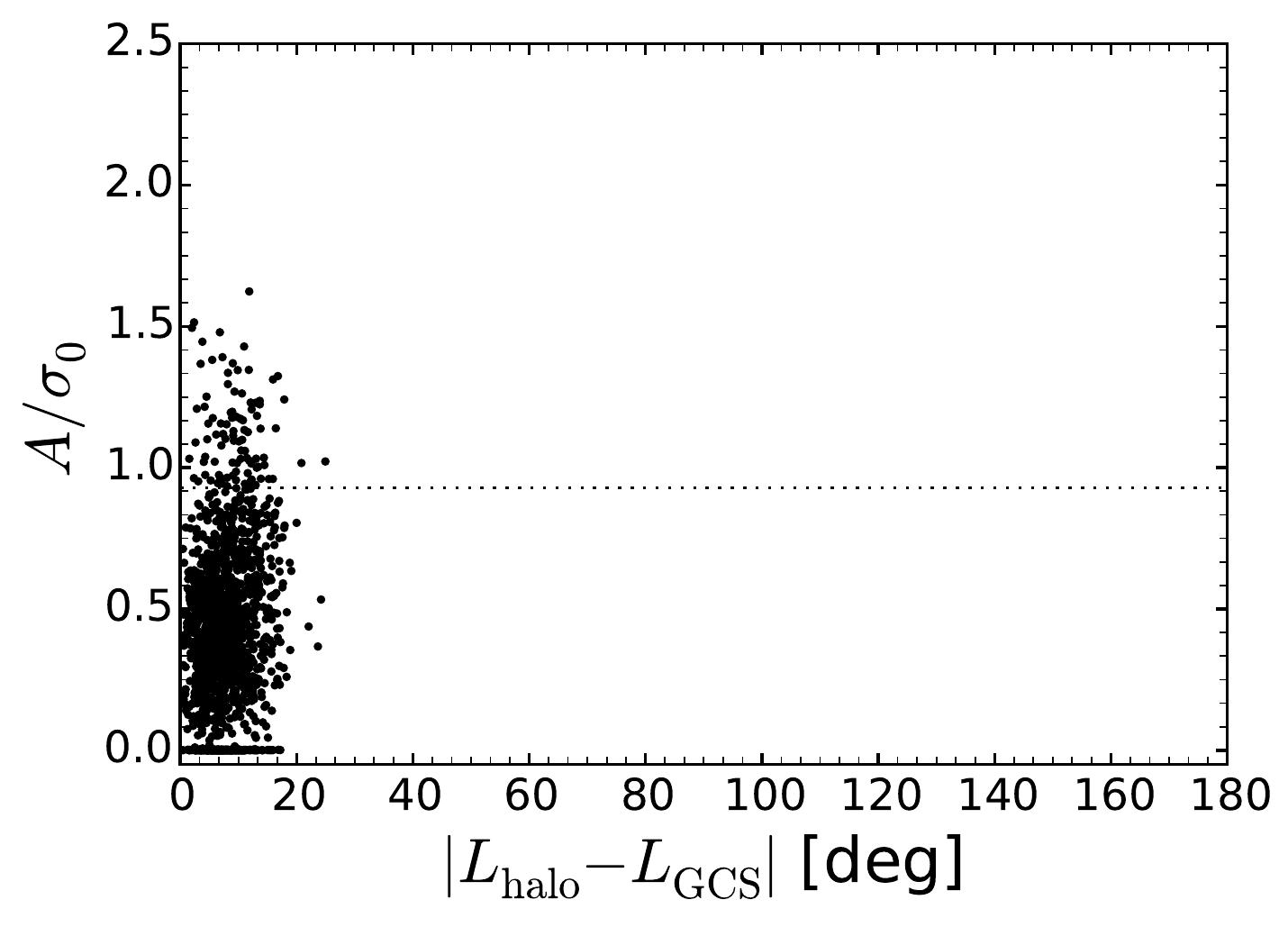} \\
\includegraphics[scale = 0.41]{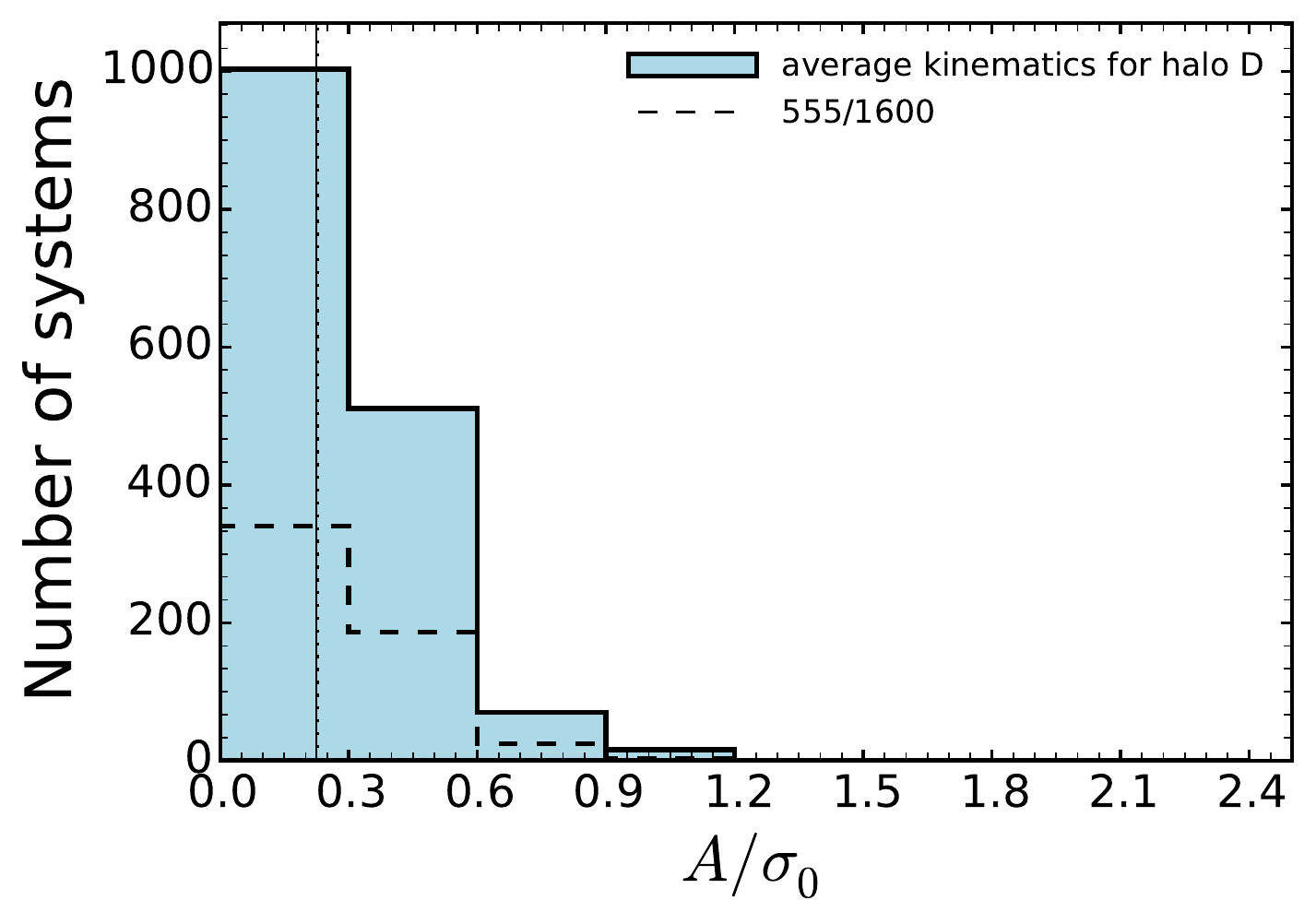} &
\includegraphics[scale = 0.41]{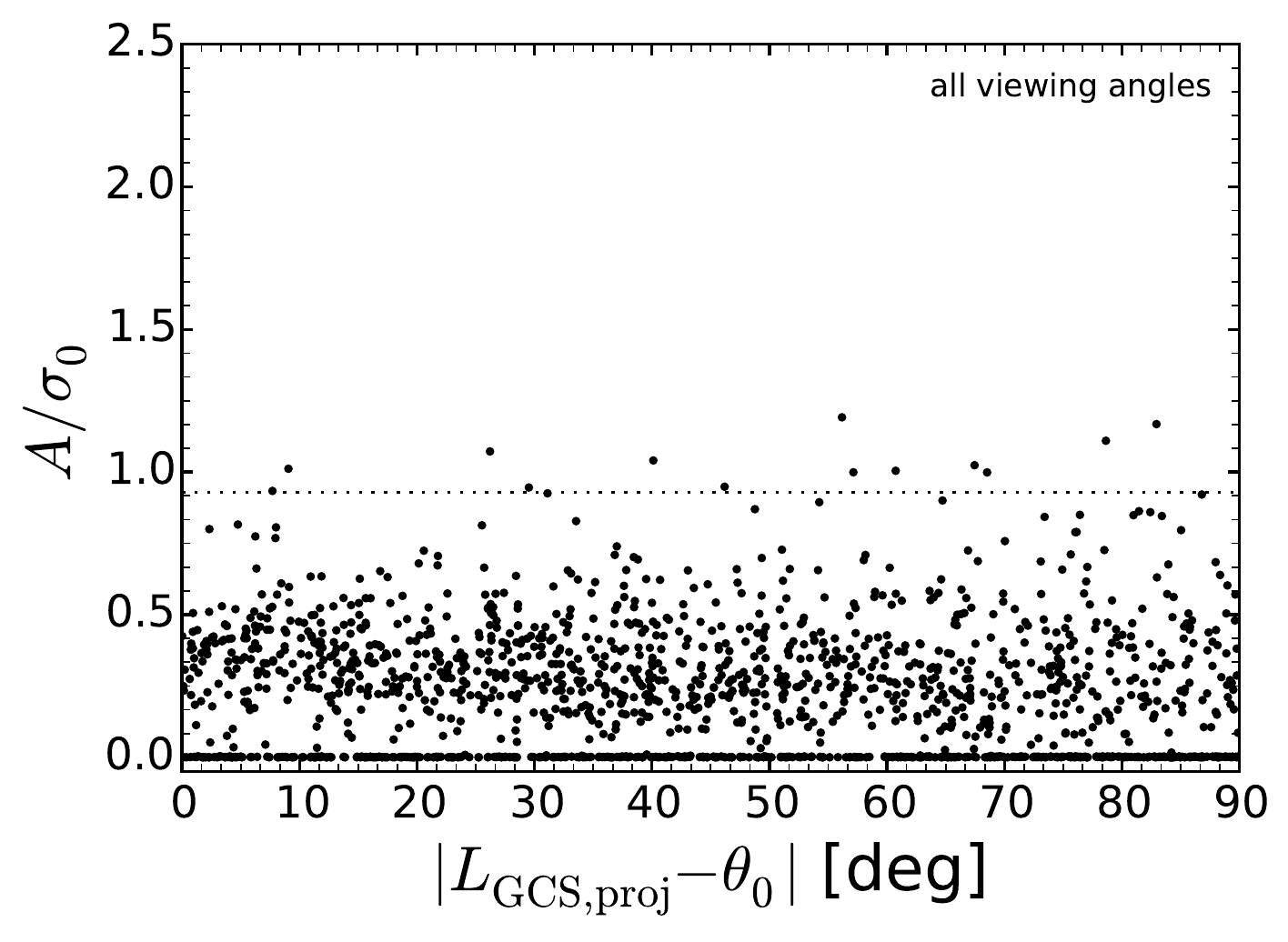} &
\includegraphics[scale = 0.41]{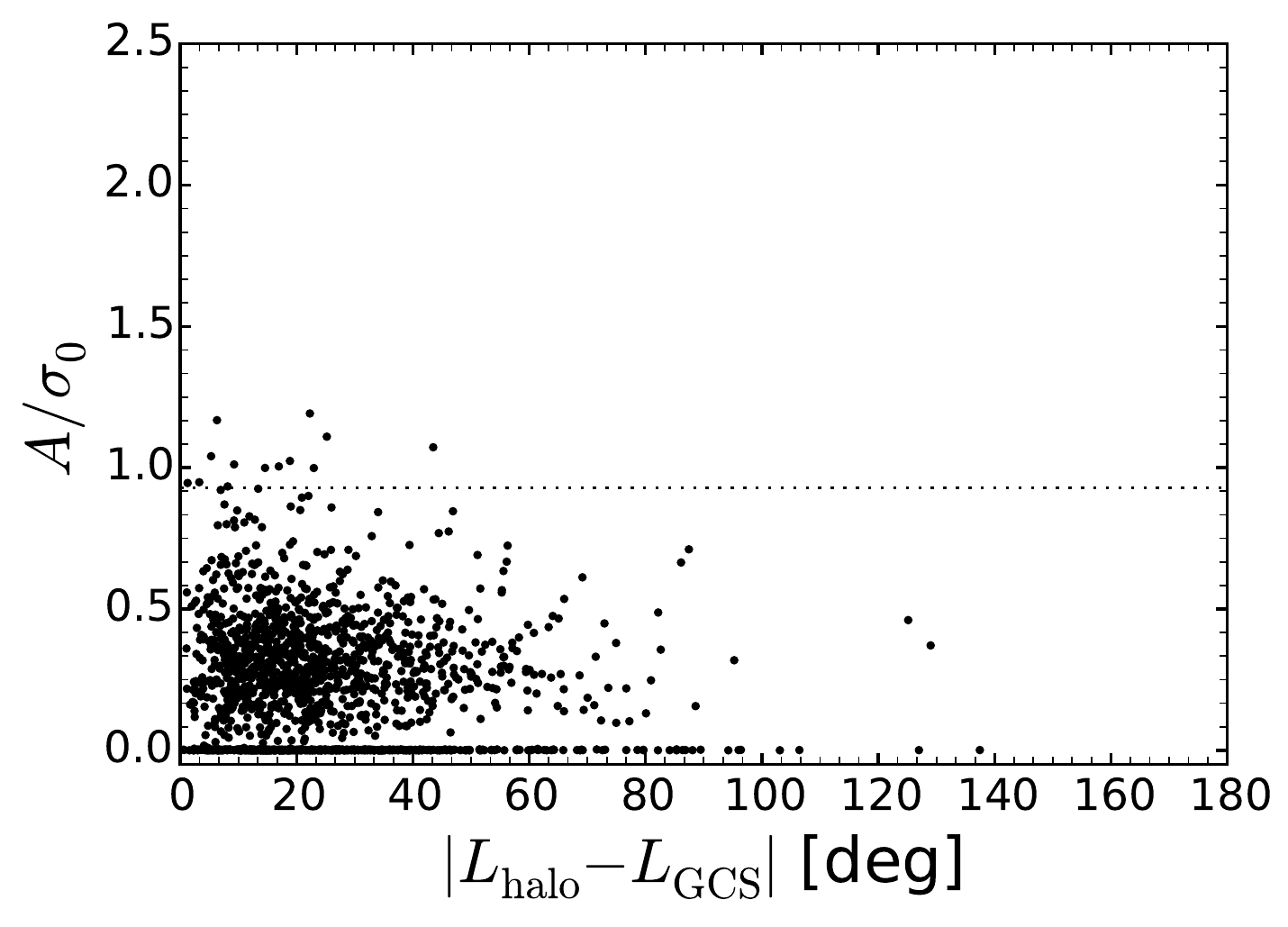} \\
\includegraphics[scale = 0.41]{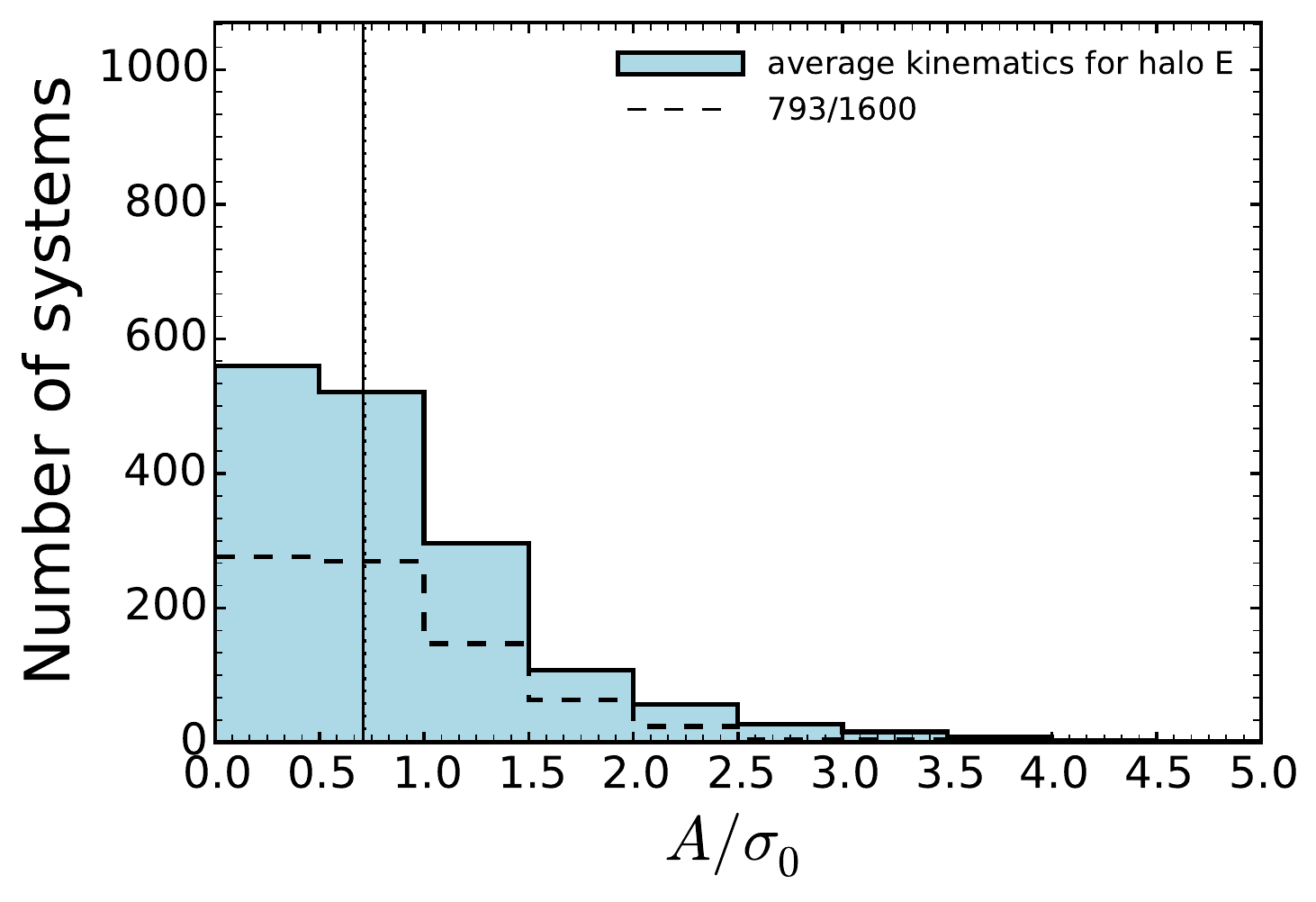} &
\includegraphics[scale = 0.41]{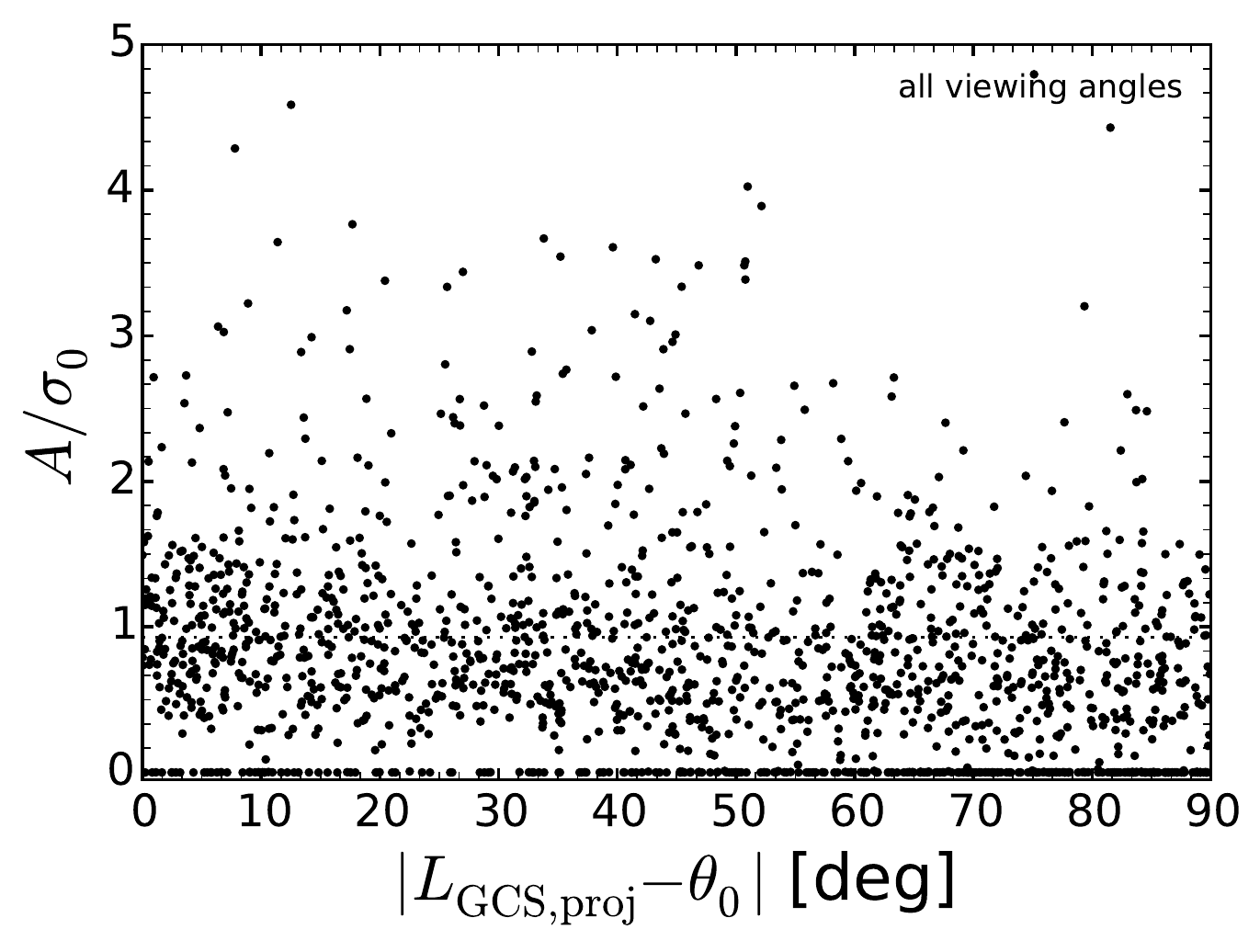} &
\includegraphics[scale = 0.41]{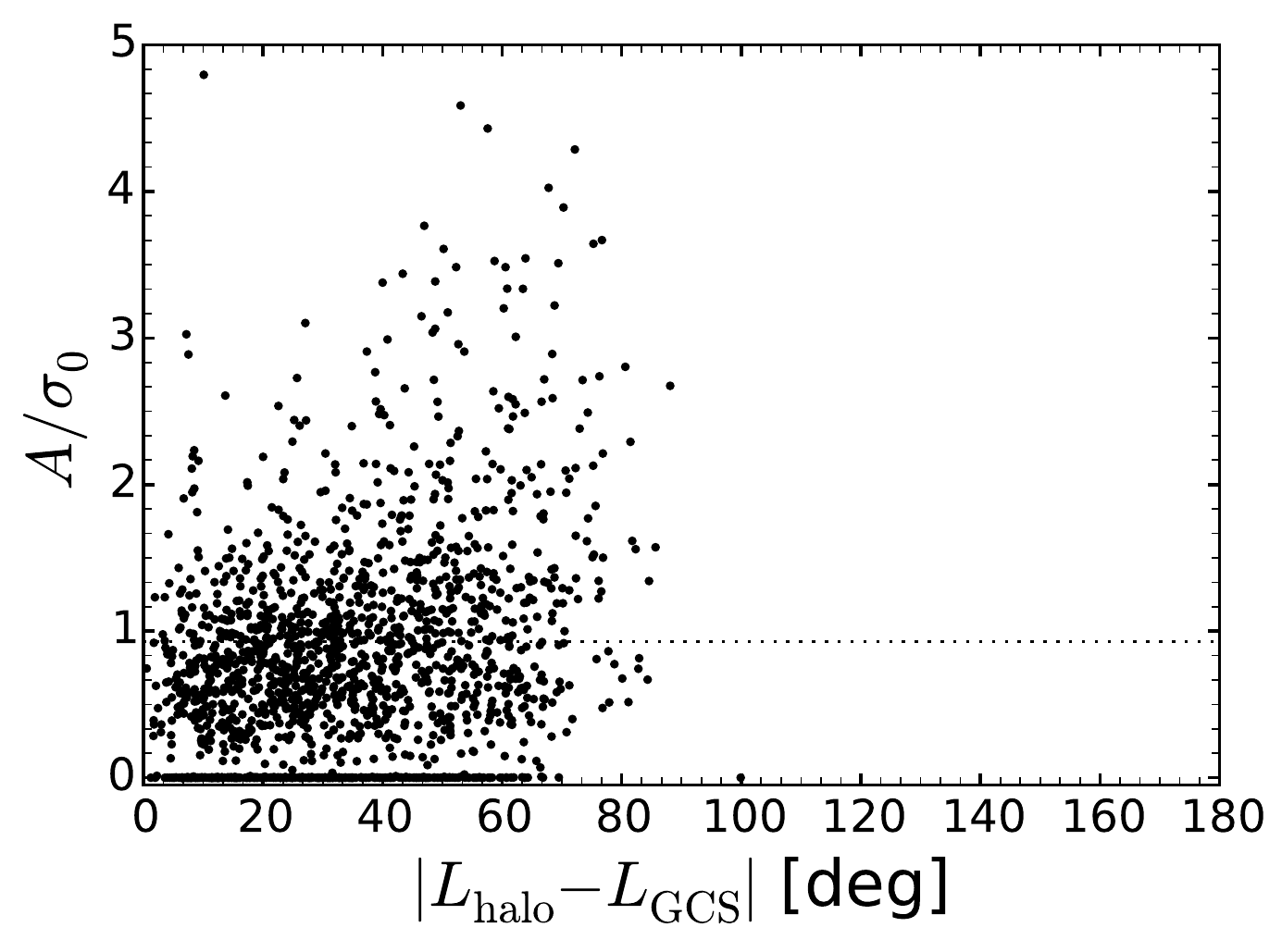}
\end{tabular}
\caption{Bulk kinematic properties of the mock GC systems  generated
in the halos around Aq-B through Aq-E. Each row corresponds to the same
Aquarius galaxy. The first column shows the distribution of kinematic ratios
for all GC systems, regardless of the viewing angle. The second column shows
the difference in the position angle between the angular momentum of the GC
systems projected on the plane of the sky and the rotation axis determined
via our analysis as a function of $A/\sigma_0$. The third column shows the
angular difference between the orientation of the 3D angular momenta
of the mock GC systems and the angular momentum of the entire stellar halo
located beyond 30~kpc in projection.}
\label{f:gridfig}
\end{figure*}

The top left panel in Figure~\ref{f:gridfig} shows that on average the mock GC 
systems around this halo show little to no rotation signal, with the mean 
kinematic ratio being less than $0.1$. Breaking down the analysis for each 
different viewing perspective (see Appendix~\ref{a:a}), we note that there is 
one observer located at $(\theta,\phi)=(90,90)$ who would see a significant mean 
value of $(A/\sigma)$ of 0.8. While this kinematic ratio is similar to the one 
around Aq-A when viewed from the favourable perspective, the circumstances that 
give rise to it are different. Closer inspection reveals that when observed from 
perspective $(\theta,\phi)=(90,90)$, the halo Aq-B is dominated by one of the 
two streams contributing $\sim 10$ clusters to its halo. This stream passes 
through the centre of Aq-B, and the GCs that it donates make up at least 50\% of 
the total number in a typical system. Thus, the entire rotation signal we are 
detecting, if present, is due to the GCs associated with this stream. The other 
and usually dominant stellar stream -- when viewed from this perspective -- 
projects on the sky in such a way  that it contributes $\sim 5$ GCs to the halo, 
but  they have little effect on the net rotation signal we are measuring.

The middle panel of the top row in Figure~\ref{f:gridfig} shows the difference
in angle between the orientation of the angular momentum, projected on
the plane of the sky, and the rotation axis determined through our kinematic
analysis as a function of $A/\sigma_0$. Even for the systems
that exhibit a significant rotation signal, there is no consistent agreement
between the true and the derived orientation of the angular momentum. This is
primarily due to the low number of clusters available for the analysis. The low
number statistics is also the primary contributor toward the large uncertainty
in determining the kinematic parameters for the systems around this halo. This
also makes it impossible to constrain the direction of the angular momentum of 
the outer stellar halo around Aq-B using the GCs as tracers, as shown in the
third panel of the first row in Figure~\ref{f:gridfig}.

\subsection{Aq-C}
\label{ss:C}

Even prior to the scaling, Aq-C had a similar virial mass to Aq-A. Aq-C has the
brightest stellar halo beyond 30~kpc amongst the galaxies comprising the 
Aquarius suite of simulations. There are nine streams in its outer halo that are
eligible to carry GCs, with five streams donating on average more than 10 
clusters. Of these five streams, two typically donate over 20 clusters, while 
one stream donates over 30 members to the GC ensembles generated around Aq-C. 
A typical mock GC system in the outer halo this galaxy contains 106 members. 
The mean uncertainty of the kinematic parameters is 
15~\kms, 31~deg, and 7~\kms\, for the amplitude, rotation axis, 
and velocity dispersion, respectively.

On average, the mock GC systems generated around Aq-C show a weak rotation
signal when observed from a random location, as shown in Figure~\ref{f:gridfig}.
With a $A/\sigma_0 = 0.4$, the GC systems around this galaxy have the highest
mean kinematic ratio when the observing perspective is not taken into account,
save for Aq-E. Even so, the probability of observing a GC system that has a
kinematic ratio at least as high as the one seen for the outer halo cluster
system of M31 is only $0.1$.

Examining the distributions of $A/\sigma_0$ for the mock cluster
systems from each of the 16 different viewing perspectives, we can
single out one such view point, $(\theta,\phi)=(90,45)$, from which
the observer detects a significant degree of rotation. In fact, when
the halo is viewed from this perspective, the $A/\sigma_0 \sim 1$. In
addition, the probability of observing a GC system with the same or
larger kinematic ratio compared to M31 is 0.6.

Figure~\ref{f:gridfig} also shows the comparison between the orientation of the
recovered rotation axis and the angular momentum of the GC systems projected
on the plane of the sky as a function of $A/\sigma_0$. These quantities agree to
within $\sim 65$~deg for 86\% of the mock GC systems generated around Aq-C.
It is interesting to see that in systems having $(A/\sigma_0 \geq 0.93)$, we
observe a relatively high misalignment of $\sim 40$~deg between these two
quantities. Even though this misalignment is not significant considering the
uncertainty with which the rotation axis is determined, it is higher than
naively expected. By carefully examining the stellar streams around Aq-C, we
found this effect to be largely caused by the most massive, dominant constituent
of this halo. This stream loops a few times around the centre of Aq-C, its shape
resembling an incomplete Rosetta pattern. While this object is the primary
driver of the rotation signal, it only contributes $\sim 25$\% of the total
number of clusters in a typical system when the halo is observed from this
perspective. The scatter in position and velocity of the remaining GCs is the
cause, since their parent streams are not aligned with the
dominant stream.

The direction of the 3D angular momentum of the entire stellar halo, on the
other hand, can be quite well constrained considering the angular momentum of
the mock GC system. The last panel of the second row in
Figure~\ref{f:gridfig} shows  that these quantities agree with each other
to within 15~deg for  86\% of the generated GC systems, regardless of the
kinematic ratio.

\subsection{Aq-D}
\label{ss:D}

At first glance, Aq-D looks quite similar to Aq-C: they both have luminous
extended stellar outer halos, and nearly identical virial masses. The outer
halo of Aq-D comprises 10 streams with stellar masses larger than 
$10^{6}$~\Msun. Four of these streams contribute more than 10 systems, while one 
stream contributes more than 25 GCs to the generated mock GC systems on average. 
A typical ensemble around Aq-D contains 94 clusters. The mean errors of the 
rotation amplitude, axis, and the velocity dispersion are 13~\kms, 47~deg, 
and 5~\kms, respectively.

Figure~\ref{f:gridfig} shows that mock GC systems generated around Aq-D, much 
like those around the other Aquarius halos, exhibit  very weak signs
of coherent rotation. The mean kinematic ratio of all mock GC systems
populating the outer halo of Aq-D is $\sim 0.2$. The chance of observing a GC
system that has similar kinematic properties to that inhabiting the outer halo
of M31 is $<1\%$. Unlike the other cases we have considered so far, the GC
systems around Aq-D show no notable rotation signal even when we consider the
16 viewing perspectives separately;  no viewing angle shows a significant
average kinematic ratio.

To understand the reason behind this, we need to consider the assembly history
of the structure around Aq-D. \citet{VeraCiro11} studied in detail the evolution
of the dark matter halo surrounding the Aquarius galaxies. In the case of Aq-D,
they found that until redshift of $z \approx 0.75$, most of the material infall
occurred along the primary filament this halo was situated on. After this time, 
material was fed into the halo of Aq-D mainly via a secondary
filament, oriented nearly perpendicular to the primary filament that dominates
the local large-scale structure. \citet{VeraCiro11} find that the dark matter
halo of Aq-D changes its orientation owing to the change in infall direction.
This change in infall direction is likely the main reason why no rotation signal
is detected in the mock GC systems around this halo.

Since we are unable to detect any significant net rotation signal, the 
determined rotation axis is very uncertain. As a result, it is not possible to 
constrain the direction of the projected angular momentum of the GC systems in 
the halo of this galaxy, regardless of the direction they are  observed  from, 
as seen in the middle panel of the third row  in Figure~\ref{f:gridfig}. 
Conversely, by knowing the true 3D angular momentum of the GCs it is possible to 
put reasonable constraints on the direction on the entire outer stellar halo 
surrounding Aq-D. By looking at the right panel of the third row in 
Figure~\ref{f:gridfig}, we find that the 3D angular momentum of the GC systems 
and that of the stellar halo agree within $\sim 40$~deg  86\% of the time.

\subsection{Aq-E}
\label{ss:E}

Aq-E is a quite compact object, similar to Aq-B. It contains the faintest
stellar halo beyond 30~kpc of all the Aquarius galaxies. The outer halo of
Aq-E comprises only five stellar streams eligible to donate GCs to the main
potential. One of these streams dominates the halo structure and contributes on
average 10 GCs. Thus there are, on average, only 20 clusters in a typical outer 
halo of Aq-E. The typical uncertainties of the derived kinematic parameters are 
32~\kms, 40~deg, and 15~\kms\, for the rotation amplitude, rotation axis, and 
the velocity dispersion, respectively.

As Figure~\ref{f:gridfig} shows, the cluster systems around Aq-E have
the highest kinematic ratio of the five Aquarius objects. The probability of
observing a GC in the outskirts of Aq-E having the same $A/\sigma_0$ as the halo
clusters of M31 is 0.7, significantly higher than for the previous cases we
examined. The cause behind this observation is a combination of low number
statistics in the number of contributing streams, and the dominating effect
that the most extended stream has on the kinematics of the mock GC systems. This
is seen more clearly  when examining the distributions of $A/\sigma_0$ as a 
function of viewing perspective. The kinematic ratio is the highest for 
perspective $(\theta,\phi)=(105,-80)$ for which an observer can see the 
dominating stream extending almost perpendicularly to the line of sight
(see Appendix~\ref{a:a}).

The GCs associated with the most extended stream in the Aq-E halo are the
main drivers behind the rotation signal, which is due to how they sample the
orbit of their parent object. These clusters alone are not sufficient to
constrain the direction of the projected angular momentum of the full halo
cluster system, however, mainly due to poor sampling of the streams, as seen in
the bottom middle panel in Figure~\ref{f:gridfig}. Even so, by knowing the 3D
angular momentum of the GC system around this halo, it is possible to  put broad 
constraints on the direction of the 3D angular momentum of the stellar halo. 
As the bottom right panel in Figure~\ref{f:gridfig} shows, these quantities 
agree within $\sim 60$~deg for 86\% of the mock GC systems around this halo.

\section{Conclusions}
\label{s:conclusions}

In this study we attempted to constrain the probability of finding a rotation
signal exhibited by GC systems residing in the outer halos of Milky Way-sized
galaxies, assembled in a cosmological setting. Using the Aquarius stellar
halos from \cite{Cooper10}, we carefully selected tagged particles to mark the
positions and velocities of GCs. This way we generated numerous sets of mock
halo GC systems, which were observed from various viewing angles. Our
comprehensive kinematic analysis yielded the following results:
\begin{itemize}
\item If a stellar halo possesses net angular momentum, it will be reflected in
the halo GC population as well, and can be detected as a rotation signal
exhibited by those clusters;
\item The amplitude of this rotation signal is highly dependent upon the viewing
angle. The rotation amplitude is significant when the halo is observed from a
perspective such that the true angular momentum vector of the GC system is
aligned by $\lesssim 40$~deg from the rotation axis determined via the projected
positions and line-of-sight velocities of the GCs;
\item The orientation of the true, 3D angular momentum vector of a halo GC
system is in very good agreement with that of the stellar halo, with the
angular difference between the two being $\sim 40$~deg on average. This is the
case provided the main streams in the halos considered are well sampled with
enough GCs located beyond 30~kpc in projection.
\end{itemize}

Considering these points, we can attempt to understand the origin of
the rotation signal exhibited by the halo GC systems, or more generally, the
origin of the common orientation of the angular momenta amongst the
substructures that comprise stellar halos, which then gives rise to the
rotation signal of the associated GC systems. The Aquarius simulations were done
in a cosmological setting, and    their assembly process can be followed through
time. In the final snapshot, it is readily seen that the distribution of stellar
streams around each Aquarius halo is anisotropic \citep{Helmi11}. Looking back
through the earlier snapshots of the simulations, we note that most of the 
matter flows into the Aquarius halos along the dominant filaments 
\citep{VeraCiro11}, which produces the anisotropic spatial
distribution of the building blocks and satellites. Hence, provided that most
of the material needed to assemble the primary halo was indeed accreted from a
preferred direction, namely along the dominant filament, a rotation signal of
that stellar halo is naturally expected to arise because the infalling 
components transfer their angular momentum to their new host. This picture is 
consistent with the simulations of \cite[e.g.][]{Li08,Libeskind11,Lovell11} and
\cite{Deason11} also regarding the satellite population. Hence, finding stellar
halos, and associated GC systems -- to have a non-negligible net angular
momentum is not surprising in a $\Lambda$CDM universe \citep{Libeskind15a}.

Another interesting question to consider is how the net angular
momentum of the stellar halo is related to the disk of a galaxy. While
the Aquarius simulations do not contain a disk component, they clearly
show that there is a preferential infall along filaments, implying
that the satellites, their debris, and the gas that is accreted onto
the primary system have similar angular momentum orientation. Thus, it
is not unreasonable to expect that, once settled, the gas to have the
same rotation sense as that of the stellar halo, albeit with a much
higher amplitude since it will have collapsed to the centre. This
picture also explains why the rotation axis of the outer halo GCs is
virtually indistinguishable from that of the GCs located in the disk
of M31. Therefore, the detection of the high rotation signal exhibited
by the outer halo GC system around M31 is not so surprising, especially
since we are observing this galaxy almost edge-on.


\begin{acknowledgements}
  We thank the anonymous referee for the useful comments that improved the
  manuscript. The authors acknowledge financial support from the European
  Research Council under ERC--StG grant GALACTICA--240271, NOVA, and NWO in the
  form of a Vici grant to AH. We are also grateful to the VIRGO consortium
  including Andrew Cooper, for collaborations carried out in the context of
  the Aquarius project.
\end{acknowledgements}

\bibliographystyle{aa}

\newpage
\begin{appendix}

\section{The GC systems of all Aq halos}
\label{a:a}

Here we show the distributions of kinematic ratios for different observing
perspective for halos Aq-B through Aq-E.

\begin{figure*}[h!]
\centering
\includegraphics[scale = 0.8]{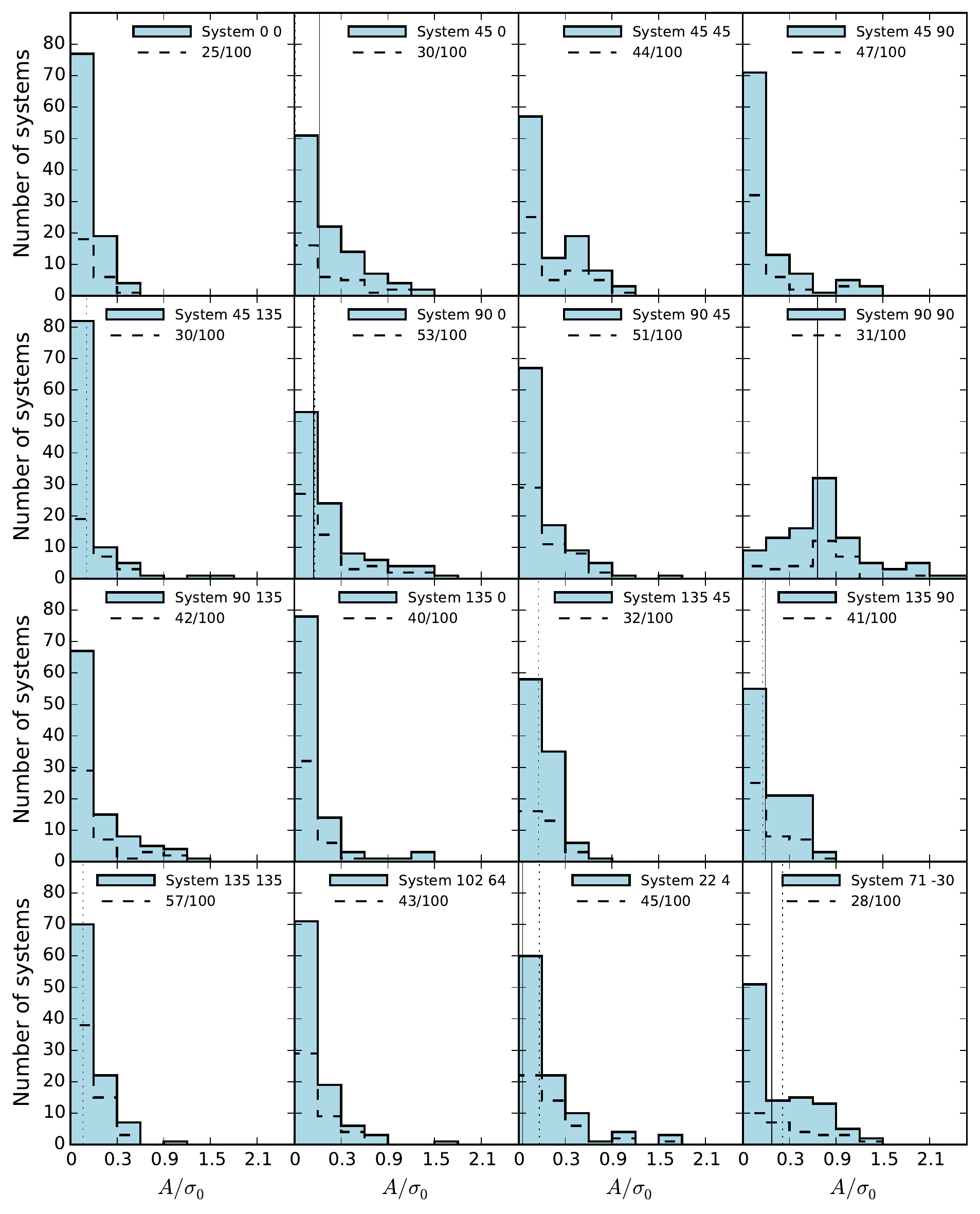}
\caption{Same as Figure~\ref{f:histfigA} in the case of Aq-B. The GC systems 
viewed from perspective $(\theta,\phi)=(90,90)$ show a significant mean 
$A/\sigma$. This is driven by GCs associated with a single dominant stream in 
the halo of Aq-B. See Section~\ref{ss:B} for details.}
\label{f:histfigB}
\end{figure*}

\begin{figure*}[h!]
\centering
\includegraphics[scale = 0.8]{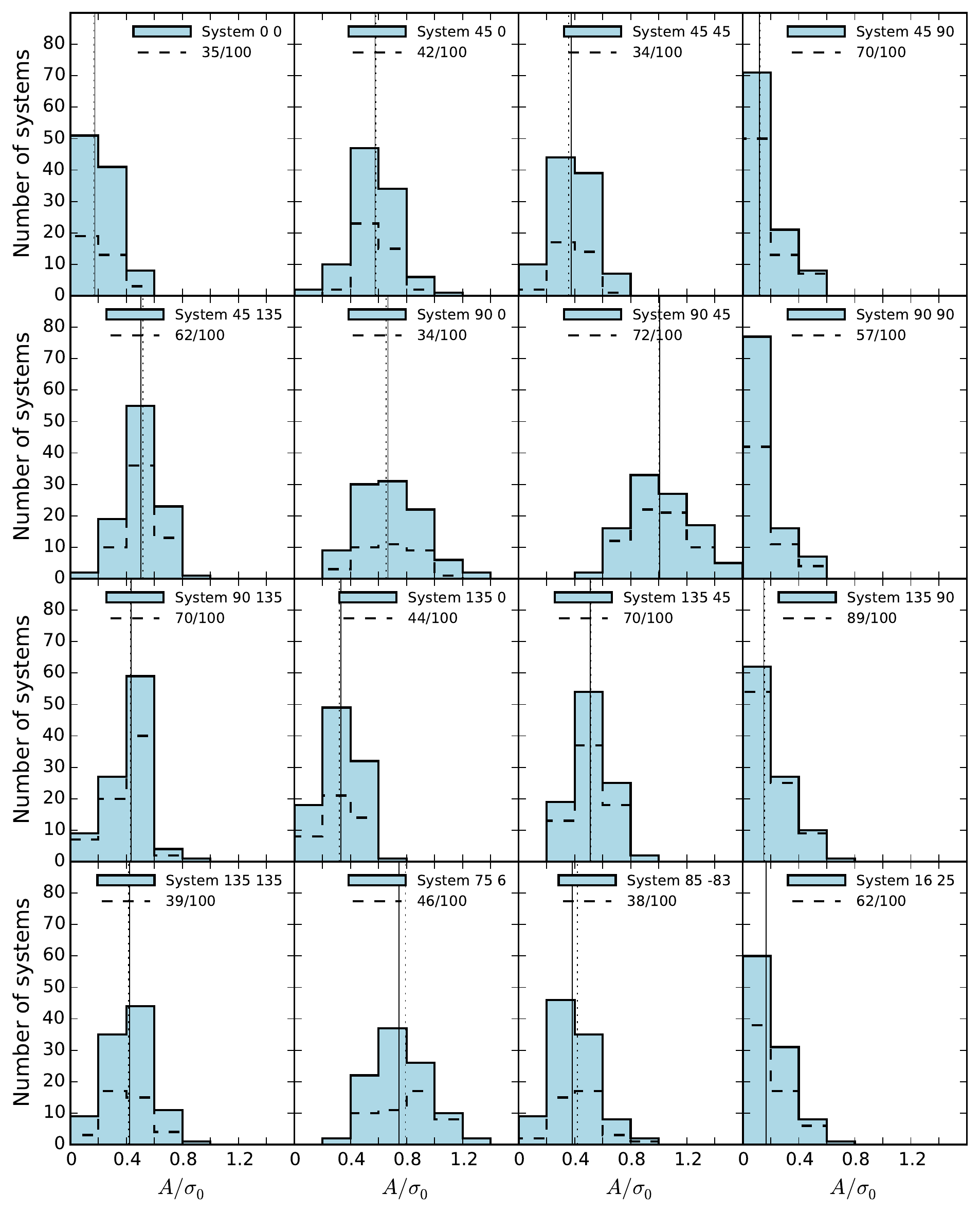}
\caption{Same as Figure~\ref{f:histfigA} in the case of Aq-C. The rotation and 
velocity dispersion have, on average, equal share in the kinematics of the GCs 
that were observed from $(\theta,\phi)=(90,45)$.}
\label{f:histfigC}
\end{figure*}

\begin{figure*}[h!]
\centering
\includegraphics[scale = 0.8]{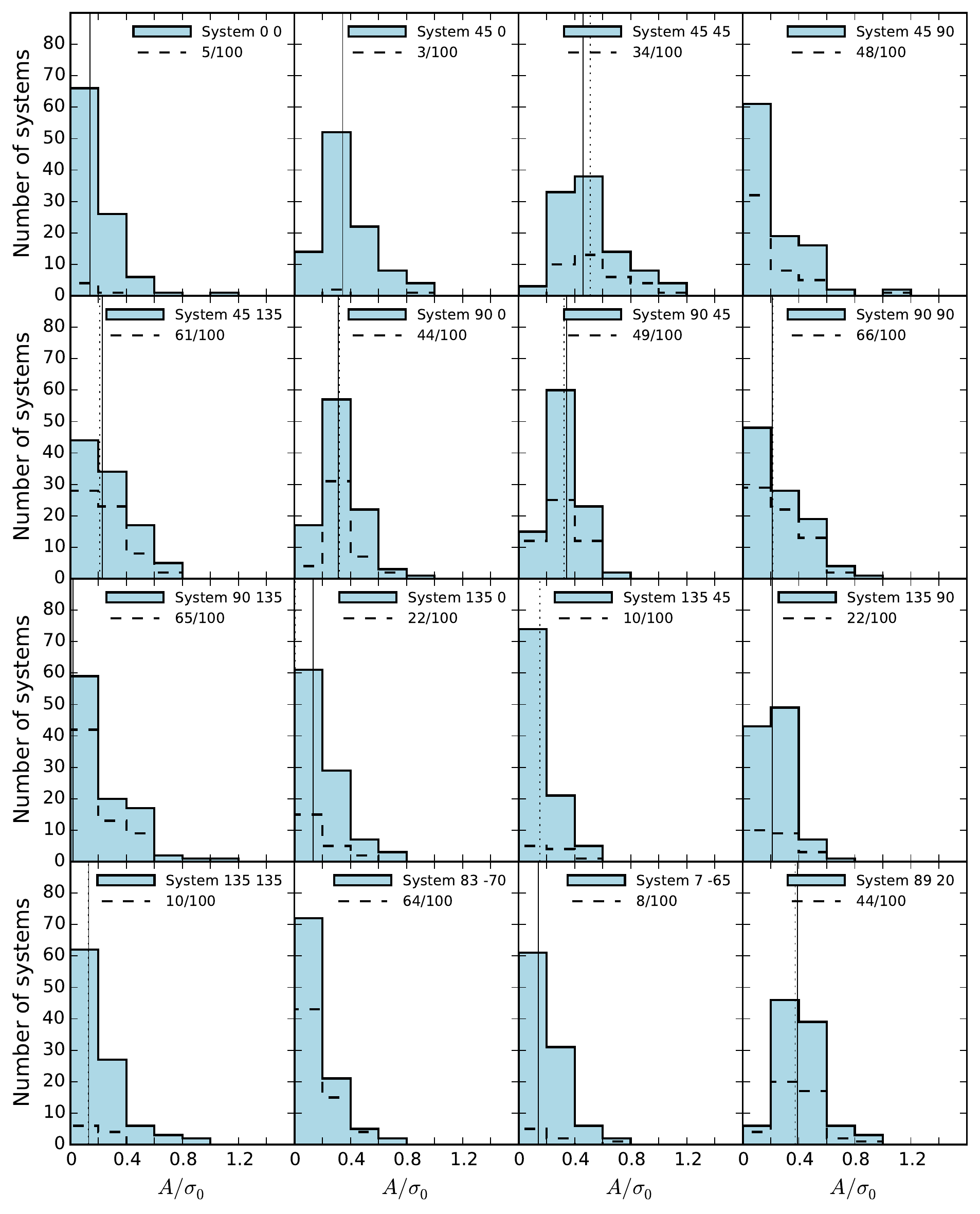}
\caption{Same as Figure~\ref{f:histfigA} in the case of Aq-D. Unlike the other
cases we  consider, the mock GCs generated around Aq-D show  very weak or no 
rotational signal regardless of the perspective they are observed from.
This is most likely related to the evolutionary history of the dark matter
halo of Aq-D. See Section~\ref{ss:D} for details.}
\label{f:histfigD}
\end{figure*}

\begin{figure*}[h!]
\centering
\includegraphics[scale = 0.8]{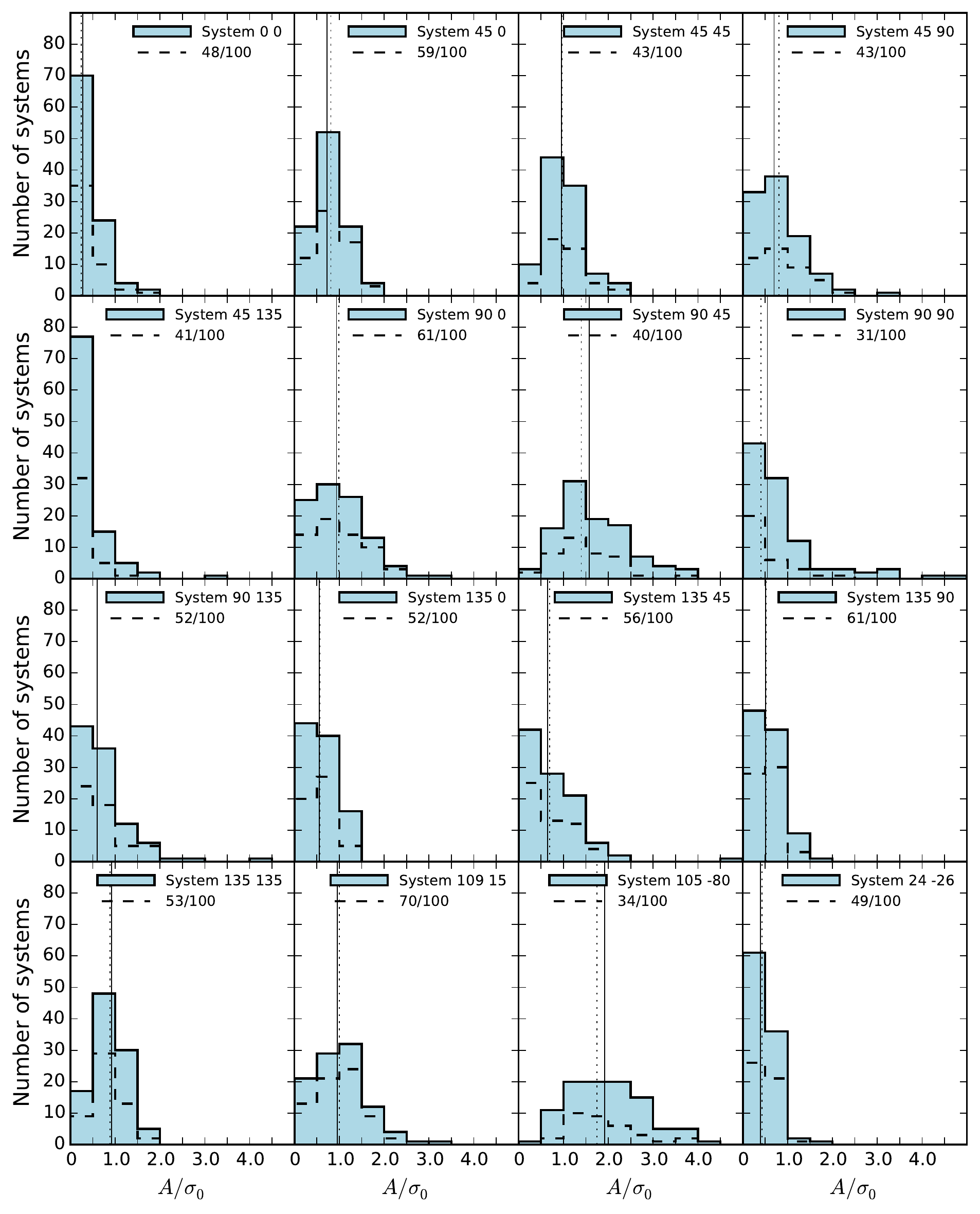}
\caption{Same as Figure~\ref{f:histfigA} in the case of Aq-E.}
\label{f:histfigE}
\end{figure*}

\end{appendix}

\end{document}